\documentclass[a4paper,fleqn,usenatbib]{mnras}
\usepackage{multirow}
\usepackage{newtxtext,newtxmath}

\usepackage[T1]{fontenc}
\usepackage{ae,aecompl}
\usepackage{aas_macros}

\usepackage{graphicx,graphics}	

\usepackage{amsmath}	

\usepackage{natbib}

\newcommand{\be}{\begin{equation}}
\newcommand{\ee}{\end{equation}} 
\newcommand{\bse}{\begin{subequations}}
\newcommand{\ese}{\end{subequations}} 
\newcommand{\bary}{\begin{eqnarray}}
\newcommand{\eary}{\end{eqnarray}}

\newcommand{\Msun}{\rm M_{\sun}}

\newcommand{\lsim}{\mathrel{\hbox{\rlap{\lower.55ex\hbox{$\sim$}} \kern-.3em\raise.4ex\hbox{$<$}}}}
\newcommand{\gsim}{\mathrel{\hbox{\rlap{\lower.55ex\hbox{$\sim$}} \kern-.3em\raise.4ex\hbox{$>$}}}}

\newcommand{\REF}{Ref-L100N1504}

\newcommand{\RECAL}{Recal-L025N0752 }

\newcommand{\sohge}{\hbox{$\rm \nabla_{(O/H)}$}}

\newcommand{\fgas}{$\rm f_{gas}$}
\newcommand{\mstar}{ {\rm M_{\ast}}}
\newcommand{\dexkpc}{dex~kpc$^{-1}$}
\newcommand{\taumm}{$\rm \tau_{MM}$}
\newcommand{\taustar}{$\tau_{25}$}
\newcommand{\Reff}{$R_{\rm eff}$}
\newcommand{\grad}{\hbox{$\nabla_{{\rm (O/H)}}$}}

\newcommand{\eagle}{{\sc eagle}}
\newcommand{\subfind}{{\sc subfind}}

\title[Evolution of metallicity gradients]{The evolution of the oxygen abundance gradients in  star-forming galaxies in the EAGLE  simulations}
\author[ Tissera et al.]{Patricia B. Tissera$^{1,2}$\thanks{E-mail:
patricia.tissera@uc.cl}\thanks{Affiliated member of ARC Centre of Excellence for All Sky Astrophysics in 3 Dimensions (ASTRO 3D).}, 
Yetli Rosas-Guevara$^{3}$, Emanuel Sillero$^{4}$, Susana
E. Pedrosa$^{5}$, 
\newauthor
Tom Theuns $^{6}$, Lucas Bignone$^{5}$\\
$^{1}$Institute of Astronomy, Pontificia Universidad Cat\'olica de
Chile, Santiago, Chile.\\ 
$^{2}$Centro de Astro-Ingenier\'ia, Pontificia Universidad Cat\'olica de
Chile, Santiago, Chile.\\
$^{3}$Donostia International Physics Centre (DIPC), Paseo Manuel de Lardizabal 4, 20018 Donostia-San Sebastian, Spain.\\
$^{4}$Instituto de Astronom\'ia Teorica y Experimental, CONICET-UNC, Laprida 960, C\'ordoba, Argentina.\\
$^{5}$Instituto de Astronom\'ia y F\'isica del Espacio, CONICET-UBA, Casilla de Correos 67, Suc. 28, 1428, Buenos Aires, Argentina.\\
$^{6} $Institute for Computational Cosmology, Physics Department, University of Durham , South Road, Durham DH1 3LE, UK.}
\date{Accepted XXX. Received YYY; in original form ZZZ}

\date{Accepted XXX. Received YYY; in original form ZZZ}

\pubyear{2020}

\begin{document}
\maketitle

\begin{abstract}
We analyse the evolution of the oxygen abundance gradient of star-forming galaxies
with stellar mass $\mstar \geq 10^{9}\rm M_{\sun}$ in the \eagle\ simulation over the redshift range $z=[0, 2.5]$. We find that the median metallicity gradient of the simulated galaxies is close to zero at all $z$, whereas the scatter around the median increases with $z$. The metallicity gradients of individual galaxies can evolve from strong to weak and vice-versa, since mostly low-metallicity gas accretes  onto the galaxy, resulting in enhanced star formation
and ejection of metal enriched gas by energy feedback. Such episodes of enhanced accretion, mainly dominated by major mergers,  are  more common at higher $z$, and hence  contribute to increasing the diversity of gradients.
For galaxies with negative metallicity gradients, we find a redshift evolution of $\rm \sim -0.03~dex~kpc^{-1}/\delta z$. A positive mass dependence is found at $z\leq 0.5$, which becomes slightly stronger for higher redshifts and,  mainly, for $\mstar < 10^{9.5} \rm M_{\sun}$.
Only galaxies with negative metallicity gradients define a correlation with galaxy size, consistent with an inside-out formation scenario. Our findings suggest that major mergers and/or significant gas accretion  can drive strong negative or positive metallicity gradients. The first ones are preferentially associated with disc-dominated galaxies, and the second ones with dispersion-dominated systems.  The comparison with forthcoming observations at high redshift will allow a better understanding of the potential role of metallicity gradients as a chemical probe of galaxy formation.

\end{abstract}


\section{Introduction} 
Chemical abundance patterns in galaxies encode information on how they formed and evolved across time, as originally pointed-out by  \citet{tinsley1980}.
 These patterns are the outcome of the complex interplay between  different physical processes, which include i) the rate at which the galaxy accretes gas from its surroundings and forms stars, ii) the rate at which elements are synthesised in stars and ejected into the interstellar medium (ISM) as stars evolve and die,
iii) the rate of mass loss from the galaxy through feedback from stars and active galactic nuclei (AGN), and iv)  stellar radial migration, among others \citep[see e.g.][]{maiolino2019}. We  can attempt to better understand how these processes operate in galaxies by characterising the spatial coherence and evolution of an abundance pattern - for example, by measuring abundance gradients across cosmic time - and comparing these observations to simulations which include these processes. Such a detailed view of the observed evolution of galaxies also provides constraints for galaxy formation models. It is topical to perform such an analysis because recent surveys have vastly improved observational constraints on abundances and simulations of galaxies have also improved in realism.

Metallicity gradients, \sohge, in more massive isolated spiral galaxies tend to be shallower than in lower mass galaxies, at least in the Local Universe (redshifts $z\le 0.1$), with the latter exhibiting a larger range in metallicity slope \citep[e.g.][]{lequeux1979, zaritsky1994}. This mass dependence is reduced when using the normalised gradients. The advent of large surveys of galaxies using integral field spectroscopy such as, for example, CALIFA \citep{sanchez2013Califa}, MaNGA \citep{bundy2015} and SAMI \citep{poet2018}, have allowed much more detailed characterisation of abundance patterns
in local galaxies \citep[e.g.][]{SanchezMen2018}. For example, there
is an indication of a weak anti-correlation between \grad\ and stellar mass, $\mstar$, also at the low-mass end \citep{sanchez2013Califa, sanchezmenguiano2016, belfiore2017}.
Galaxies showing signs of interactions often exhibit less negative or even positive metallicity gradients (also called 'inverted'), which contrasts with the negative gradients of non-interacting galaxies of the same stellar mass \citep[e.g.][]{kewley2006, rupke2010}. Clearly, there is a lot to learn about how galaxies evolve from this new data.

Metallicity gradients exhibit a relatively large scatter between galaxies at higher redshifts, $z>0.1$, \citep[e.g.][]{cresci2010, queyrel2012, stott2014, carton2018}, with a significant fraction of galaxies with shallow or positive gradients\footnote{The accurate determination of abundances is challenging, with measurements often resulting in
systematic offsets between different metallicity indicators \citep[e.g.][]{kewley2010}, a dependence on the spatial resolution of  observations \citep[e.g.][]{yuan2013} and biases due to sample selection, amongst other observational challenges.}.
\citet{curti2019} reported metallicity gradients for 42 lensed galaxies with $z\in[1.2, 2.5]$, concluding that $\sim 89$\% of the studied galaxies have metallicity gradients shallower than 0.05 \dexkpc. \citet{wang2019b} analysed 79 lensed galaxies in the redshift range of $z\in [1.3, 2.3]$, concluding that lower-mass galaxies have steeper negative gradients,
$\grad \approx -0.020 \pm 0.007 \log(M_\ast/10^{9.4}\Msun$)$~{\rm dex~kpc}^{-1}$.

A correlation between metallicity and star formation activity in galaxies  is expected as, on the one hand, metals  are an important coolant in the ISM of galaxies and hence likely affect both the rate at which stars form and the efficiency of stellar feedback. On the other hand, physical processes, such as gas inflows and mergers, that modify the properties of the ISM will also disturb  the chemical abundances distributions.
Indeed, \citet{stott2014} and 
\citet{curti2019} find that galaxies at $z\sim 1$ with high specific star formation rate, $\dot M_\ast/M_\ast$ (hereafter sSFR), tend to have shallow or positive metallicity gradients. They suggest that this trend is caused by the action of mergers, which can simultaneously dilute the metallicity of the ISM while increasing the  amount of gas available for feeding the star formation activity. With such gas-rich mergers more frequent at high $z$, we may expect that the scatter in gradients increases with $z$. As also pointed out by these authors, these observations are likely to be affected by sample bias (see also \citealt{poet2018, curti2019}) since a high star-forming galaxy is, of course, much easier to detect at higher $z$ than a galaxy with low SFR. In addition, the decrease in spatial resolution towards higher $z$ may artificially flatten any intrinsic gradient \citep[e.g.][]{curti2019}. Future observations will hopefully mitigate against these biases, by improving sample selection and spatial resolution \citep[e.g.][]{maiolino2019}.

It remains relatively poorly understood which physical processes cause the large scatter in metallicity gradients,  modulate their evolution with redshift and, hence, how to unveil them from abundance patterns.
Earlier theoretical models suggested that the inside-out formation of a galaxy will result in negative metallicity gradients \citep[e.g.][]{tinsley1978, prantzos2000}. 
More recent models of abundance evolution attempt to account for the inflow of pristine gas, and the outflow and recycling of enriched gas \citep[e.g.][]{molla2017} as well as other processes \citep[e.g.][]{chiappini1997, belfiore2019, sharda2021}, finding similar results if a suitable balance among these processes is attained. 

Cosmological simulations that include chemical evolution \citep[e.g.][]{mosconi2001, lia2002, koba2007} are able to reproduce galaxies with metallicity gradients
in general agreement with observations, reporting negative metallicity gradients in galaxies which formed inside-out \citep[e.g][]{pilkington2012, tissera2016a,ma2017,hemler2020}.
Stellar feedback is found to play a crucial role in modulating the metallicity distributions via the regulation of the star formation activity and the triggering of mass-loaded galactic winds that can transport material outside galaxies
\citep{gibson2013, tissera2019}.   
 Additionally,  other physical processes such as galactic fountain \citep[e.g.][]{grand2019}, tidally exchanged material  \citep{rupke2010}, bar formation \citep[e.g.][]{fragkoudi2020}, cold gas inflows to the central regions \citep[e.g.][]{ceverino2016,collacchioni2019} and
low-metallicity gas inflows triggered during galaxy-galaxy interactions \citep[e.g.][]{bh96,tissera2000, perez2011, dimatteo2013,  perez2006,moreno2019, bustamante2018}.
The combined effects of these physical processes modulate the evolution of the metallicity gradients as a function of redshift. \citet{tissera2016} showed that galaxies with a close companion were more likely to have positive gradients and that the evolution of the metallicity gradients was stronger for galaxies with $\mstar \leq 10^{10} \rm \Msun$ than in more massive systems. \citet{ma2017grad} reported a significant evolution of metallicity gradients on timescales of $\leq 1$Gyr, with strong gradients associated with starbursts in the {\sc FIRE} simulations. \citet{hemler2020} determined that galaxies in TNG50 grow more negative with redshift at a rate of $\rm -0.02 ~dex~kpc^{-1}/\delta z$. The differences found in the evolution of the metallicity gradients between these works can be interpreted as consequence of the different energy feedback models adopted.

In this paper, we analyse metallicity gradients of galaxies identified in the \eagle\ suite of simulations described by \cite{schaye2015}, concentrating in particular on the highest-resolution \RECAL\ simulation that has a baryonic mass resolution of $\sim 10^5M_\odot$.
We measure the metallicity gradient in the star-forming gas in  simulated  galaxies over a relatively extended redshift range, $z\in[0,2.5]$. The \eagle\ simulation uses subgrid models to account for physics below the resolution scale, with parameters calibrated to reproduce the $z=0$ galaxy stellar mass function, galaxy sizes and stellar mass - black hole mass relation, as described by \cite{crain2015}. The simulations reproduce several key observables that were not part of the calibration strategy, including
the gas content of galaxies at a given mass \citep{lagos2017, crain2017}, the evolution of the galaxy stellar mass function and galaxy sizes \citep{furlong2015, furlong2017}, 
the bi-modality of galaxy colours \citep{trayford2015, trayford2017}, the mass-metallicity relation \citep{derossi2017}, the scale-resolved metallicity-star formation relation \citep{trayford2019b}, the azimuthal variation of the metallicity gradients at $z=0$ \citep{solar2020},  the central metallicity gradients \citep[within 0.5 the half-mass radius][]{collacchioni2019} and non-parametric morphologies \citep{bignone2020}. Most closely related to the current paper, \citet[][hereafter T19]{tissera2019} showed that the metallicity gradients of star-forming regions in the disc components of central galaxies of the \REF~ simulation fall within the observed scatter
at $z\sim 0$. The median gradient is close to zero with little dependence on galaxy mass, but galaxies with a more extended disc tend to have a shallower gradient. T19 claim
that a high star formation efficiency across the discs together with strong feedback
 are the main cause of the shallow  metallicity gradients in the simulation. 

 In this work,
we will compare simulated and observed metallicity gradients, in particular
the gradient of the oxygen over hydrogen abundance, adopting the abundance indicator\footnote{The abundance indicator (O/H) is defined as the ratio between the atomic number densities of the corresponding chemical elements. We note that all logarithms correspond to base 10, but for  the sake of simplicity, the sub-fix has been dropped.} $12 + \log_{10}({\rm O/H})$, commonly used for HII regions. The units of \sohge\ are dex per kpc.
We note that a  negative gradient means that the centre of the galaxy is more highly enriched (in oxygen) compared to its outskirts. A galaxy with a larger  absolute value of $|\grad|$ is said to have a  steeper gradient - irrespective of the gradients sign; if $| \grad |$ is smaller, the gradient is said to be shallower.

 We present results on the evolution of metallicity gradients in star-forming regions of simulated galaxies with a wide variety of morphologies. Our sample comprises a total of 957 central (i.e. non-satellite) galaxies with redshift $z\in [0.0,2.5]$ selected from \RECAL. We study the dependence of metallicity gradients on redshift, stellar mass, galaxy size, sSFR, and merger history.
 
The paper is organised as follows. Section \ref{sec:sims} briefly summarises the main characteristics of the simulations used, \RECAL. In Section \ref{sec:analysis}, we analyse the dependence of the metallicity gradients on stellar mass, galaxy size, sSFR and redshift. In Section \ref{sec:mergers}, we investigate the dependence of the metallicity gradients on the recent merger history, the gas fraction and the morphology of galaxies. In Section \ref{sec:individual}, we analyse the metallicity gradients of a set of galaxies along their merger trees. Our conclusions are summarised in Section~\ref{sec:conclusions}.

\section{Simulations} \label{sec:sims}
The \eagle\ project is a suite of cosmological hydrodynamical simulations\footnote{We use the database publicly available by \citet{mcAlpine2016}.}, assuming a $\Lambda$CDM universe with
cosmological parameters taken from \citet{planck2014}, $\rm \Omega_{\Lambda} = 0.693$, $\rm \Omega_{m} = 0.307$, $\rm \Omega_{b} = 0.04825$, $h = 0.6777$ ($H_{\rm 0} = 100\ h$ $\rm km\ s^{-1}\ Mpc^{-1}$), $\rm \sigma_{8} = 0.8288$, $n_{\rm s} = 0.9611$, and $Y = 0.248$, where 
the cosmological parameters are denoted with their usual symbol.

The simulations were performed with a version of the {\sc gadget-3} code \citep{springel2005} with some changes to the hydrodynamics solver and time-stepping scheme, encapsulated in
the {\sc anarchy} model described by \citet{schaller2015}. The simulations use a set of subgrid models to account for physical processes below the resolution of the simulation, such as star and black hole formation, stellar evolution, and energetic feedback from stars and accreting black hole, see \cite{schaye2015} for full details.

Relevant for this work, the simulation includes stellar evolution and models for the nucleo-synthesis and dispersal of eleven elements as described by \citet{wiersma2009}.
A \citet{chabrier2003} stellar initial mass function (IMF) was adopted, with stars in the range of $M=0.1\to 100M_\odot$. The star formation scheme stochastically transforms gas particles into stars above a threshold density that is metallicity dependent, as described by \citet{schaye2010}.  The simulation also includes radiative cooling and photo-heating models as described by \citet{wiersma2009a}. Energetic feedback from stars is implemented stochastically, as described by \cite{dallavecchia_schaye2012}, with subgrid parameters calibrated 
as described by \citet{crain2015}. 

We use the \RECAL simulation of table~3 of \cite{schaye2015}. The linear extent
of this simulation is $25$ cMpc (co-moving mega-parsecs), and it uses
$752^{\rm 3}$ initial baryonic and dark matter particles.
The  gravitational softening ($0.35$ pkpc, proper kilo-parsecs) is kept constant in proper units below $z=2.8$; at higher $z$ the softening is kept constant in co-moving units
at 1.33~ckpc. The mass resolution is $2.26 \times 10^5~\Msun$ and $1.21 \times 10^{6}~\Msun$ for the initial gas and dark matter particles, respectively \citep[see][for more details]{schaye2015}. 

Galaxies are identified in the simulation snapshots using the friends-of-friends
\citep[FoF,][]{davis1985} and \subfind\ \citep{springel2001a, dolag2009} algorithms.
We restrict the analysis to central galaxies, and hence, we do not study galaxies identified as satellites by the \subfind \ algorithm. We study both the evolution of samples of galaxies identified in a given snapshot\footnote{We use snapshots at redshifts:  0.10, 0.18, 0.26, 0.50, 0.61, 0.73, 0.86, 1.0, 1.23, 1.50, 1.73, 2.0, 2.23 and 2.5.} and the evolution of individual galaxies along their merger tree. 

\subsection{Main properties of the simulated galaxies} \label{sec:sim_glx_z}

We use the method of T19 to identify a disc and a spheroid in every galaxy, based on the angular momentum distribution of the stars and gas. Briefly, the direction of the total angular momentum, ${\bf J}$, is set by all the stars (or gas particles) of the galaxy, and we chose the $z$-axis along ${\bf J}$. Individual particles are assigned to the disc or the spheroidal component according to the value of their orbital circularity parameter, $\epsilon= J_{\rm z}(E)/J_{\rm z,max} (E)$. Here, $J_{\rm z}(E)$ is the angular momentum along the $z$-axis of the particle with binding energy $E$, and $J_{\rm z,max} (E)$ is the maximum value of $J_{\rm z}$ over all particles with the same binding energy. Particles in a disc are on a nearly circular orbit for which $J_{\rm z}$ is maximum, therefore disc particles tend to have $\epsilon\sim 1$ (or $\epsilon\sim -1$ if in a counter-rotating disc), whereas particles in the spheroid have $\epsilon\sim 0$ \citep[see ][for details on this procedure]{tissera2012,pedrosa2015}. In practise, we assign a particle to the disc if $\epsilon > 0.4$, as in T19. The bulge-disc decomposition allows the definition of the stellar mass disc-to-total ratio (D/T), which will be used as a morphological indicator (i.e. larger values of D/T implies that galaxies are more rotationally-supported). \cite{correa17} used a similar technique to classify galaxies as discs or spheroid, and showed that this classification correlated strongly with galaxy colour, with discs blue and spheroids red. This is, of course, reminiscent of the original Hubble sequence of galaxies \citep{Hubble1927}.

 For each galaxy we compute the stellar half-mass radius, $R_{\rm eff}$, the star formation rate, SFR, the specific star formation rate, sSFR, and the gas fraction $f_{\rm gas}=\rm M_{ \rm gas}/(\mstar + \rm M_{ \rm gas})$. We also compute the time, \taumm, since the last merger
using the merger tree (see Section~\ref{sec:mergers} for a detailed description). Galaxies with $R_{\rm eff} < 1$~kpc (only 3 times larger than the gravitational softening) and  stellar masses smaller than $10^9~\Msun$ (having fewer than $\sim 10^4$ star-particles and typically fewer than $\sim 10^3$ gas particles) are excluded from our sample to mitigate the effects of low numerical resolution on the determination of the metallicity profiles.

We measure \sohge\  of star-forming gas (hereafter abbreviated as SFG). Such gas satisfies all simulation requirements for being converted to stars (in terms of density, over density, and temperature thresholds); the \eagle\ star formation prescription converts such gas particles to stars stochastically as described by \cite{schaye2008}. 

 Abundance gradients of galaxies are computed following T19 as follows. The angular moment vector ${\bf J}$ of the galaxy is assumed to be perpendicular to the galaxy's disc.
We then construct cylinders parallel to ${\bf J}$  of varying radii,
and compute the SFR-weighted metallicity of SFG particles in each cylindrical shell.
We compute the radial gradient of these cylindrical shells, provided the galaxy contains more than 100 SFG particles in the radial range $\in [0.5, 1.5]R_{\rm eff}$. 
This radial interval is similar to that adopted in the MaNGA and CALIFA observational surveys. Appendix~\ref{append:grad} contains more information on the procedure
for computing gradients.

We also computed the \sohge\ for all SFG, {\em i.e.} not restricting the SGF to be in a disc. In both cases, the linear fits were performed within the radial range [0.5, 1.5]$R_{\rm eff}$. These two measures of the gradient correlate very tightly, as can be seen from Fig.~\ref{gradgradgas}. The good correspondence between the two gradients can be assessed from the Spearman coefficients included in this figure. The good correlations detected at all stellar masses indicates that most of the SFG is located in a rotational-supported disc \citep[see also][]{trayford2019a, rosito2019}, which is, of course, expected. In what follows, we will use the \sohge\  estimated from the SFG in the disc.

Finally, galaxies become increasingly irregular in shape with increasing redshift, both observationally and in the simulations. The value of the metallicity gradient \sohge, computed as described above, only captures approximately any gradients in abundance in what is typically a clumpy and irregular ISM.  A similar issue affects observations \citep{maiolino2019}. This is particularly important at $z > 2$.

\section{The statistics of metallicity gradients} \label{sec:analysis}

To carry out this analysis, we bin the sample of simulated galaxies in stellar mass and redshift. The stellar mass bins are $[10^9, 10^{9.5}]~\Msun$, ($10^{9.5}, 10^{9.9}]~\Msun$ and larger than $10^{9.9}~\Msun$; for reference, the most massive galaxy in the simulation has $\rm M_\ast=10^{10.8}~\Msun$. The redshift bins are $[0,0.5]$, $(0.5,1.5]$ and $(1.5,2.5]$.  The number of galaxies in the sample decreases rapidly with increasing $z$, partly because the simulated volume is relatively small and massive galaxies are rare at high $z$, but additionally because the simulated galaxies are increasingly poorly resolved with sizes not much larger than the gravitational softening. We  decided against including them  in the statistical analysis presented below as already mentioned in the previous section.

\subsection{Evolution of metallicity gradients}

\begin{figure}
\resizebox{8cm}{!}{\includegraphics{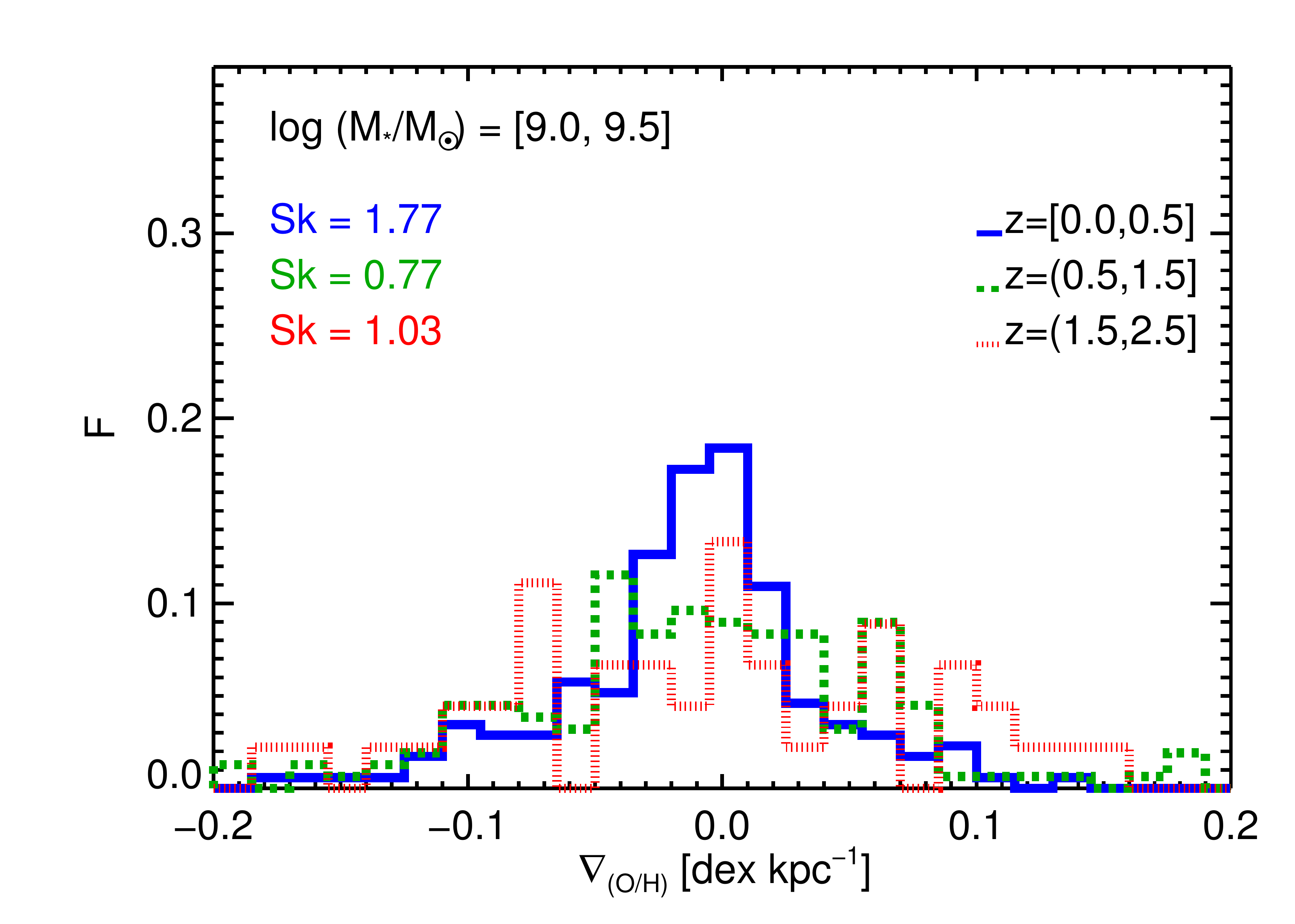}}\\
\resizebox{8cm}{!}{\includegraphics{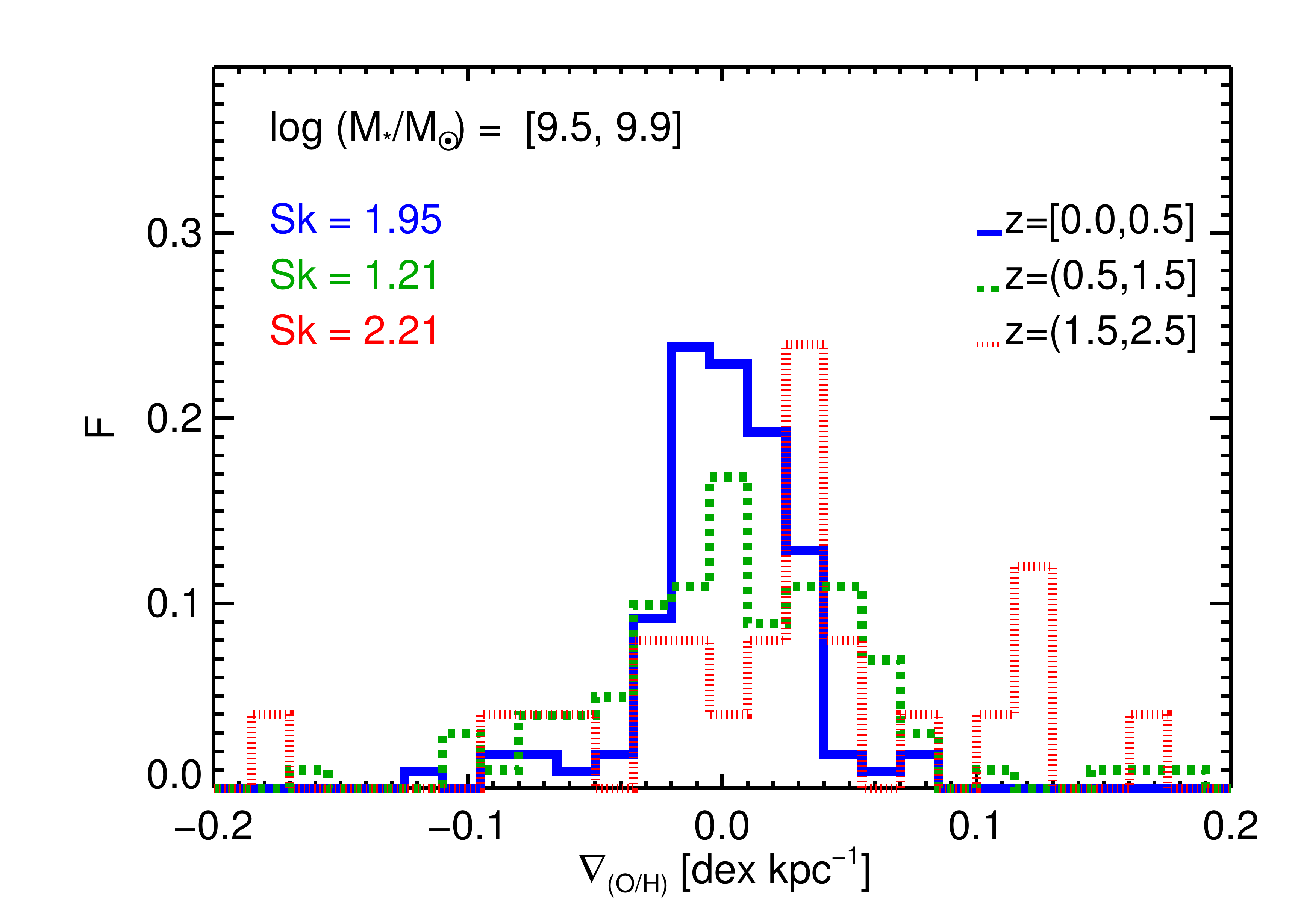}}\\
\resizebox{8cm}{!}{\includegraphics[trim={0 40 0 0}]{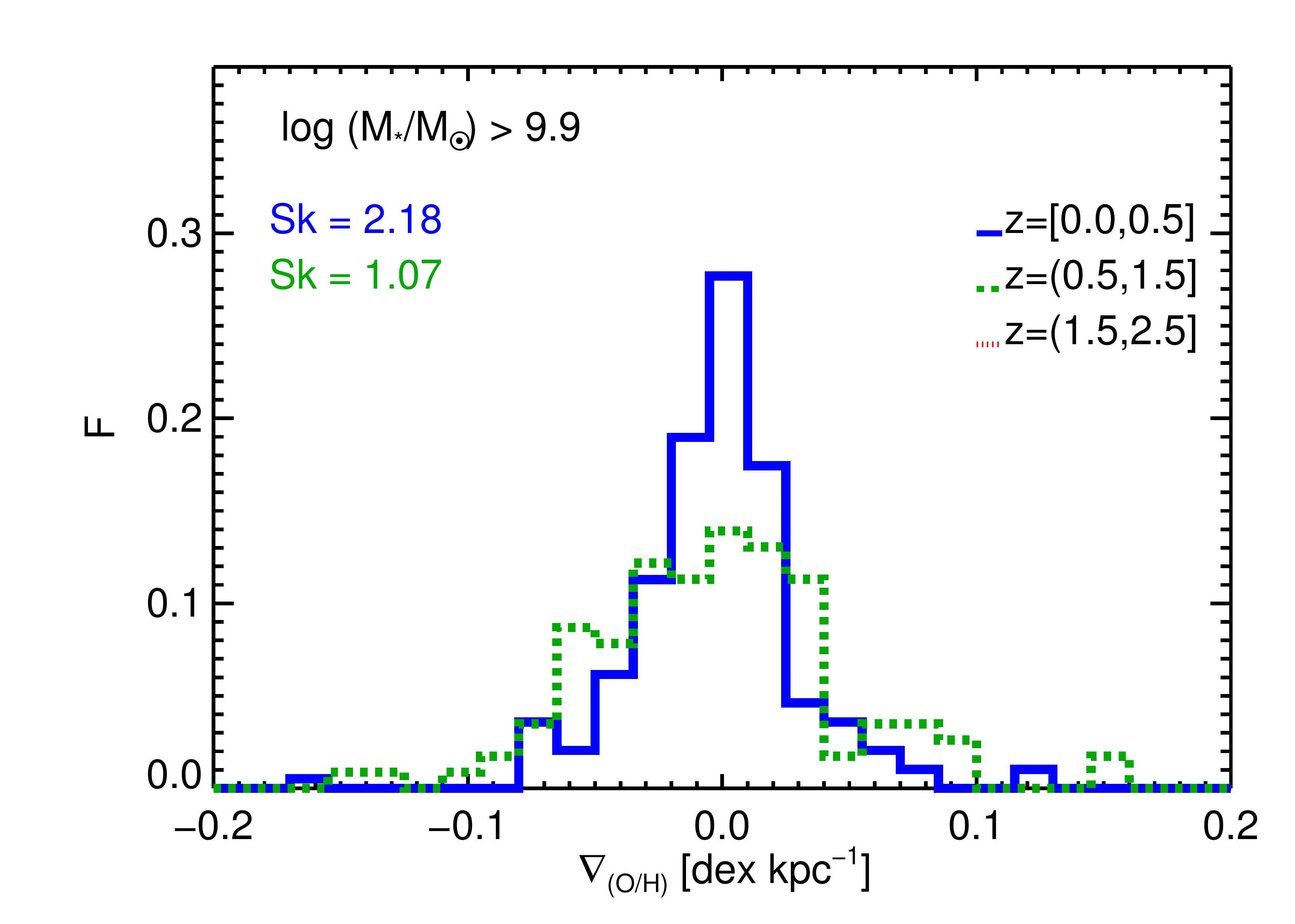}}\\
\caption{Histograms of metallicity gradient, \sohge, for galaxies from the \RECAL\ \eagle\ simulation, binned by redshift (colours) and stellar mass (upper to lower panels). 
Each histogram is normalised to unity. The results for the highest mass bin are not shown for the highest redshift interval due to the small number of simulated galaxies in that bin. The median value for each histogram is close to zero, with outliers up to $\pm 0.2~{\rm dex~kpc}^{-1}$. The distribution is skewed toward positive gradients as quantified by the skewness, $\rm Sk$, indicated in the panels.}
\label{histos}
\end{figure}

\begin{figure*}
\resizebox{18cm}{!}{\includegraphics[trim={0 40 0 0}]{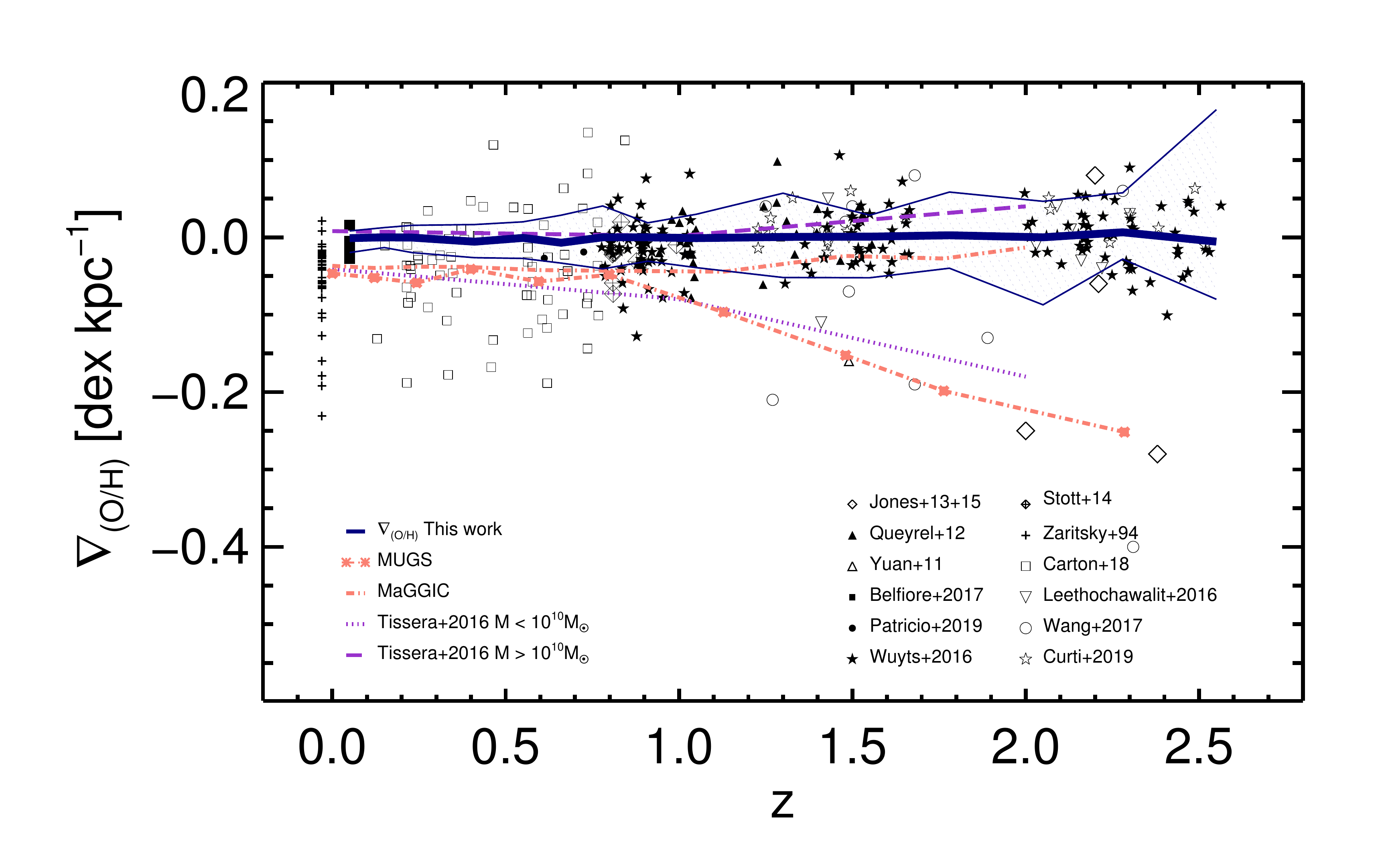}}
\caption{Evolution of the metallicity gradient, \sohge. The {\em dark blue line}
is the evolution of the median of \sohge\ of the star-forming gas in the \RECAL\ \eagle\ simulation, with faint blue shading includes the 25-$75^{\rm th}$ percentile range. Observational
values of the gradient are shown as black symbols, taken from 
\citet[][crosses at $z=0$]{zaritsky1994}, 
\citet[][open triangles]{yuan2011},  
\citet[][solid circles]{queyrel2012},
\citet[][open diamonds]{jones2013, jones2015}, 
\citet[][crossed diamonds]{stott2014}, 
\citet[][solid stars]{wuyts2016},
\citet[][inverted triangles]{lee2016}, 
\citet[][open circles]{wang2017}, 
\citet[][small solid squares displaced to $z=0.1$ for clarity represent median values of gradients ]{belfiore2017},
\citet[][open squares]{carton2018}, 
\citet[][solid circles]{patricio2019} and 
\citet[][open stars]{curti2019}. 
 Observational values have been re-scaled to a \protect\cite{chabrier2003}  IMF and the \eagle\ values of the cosmological parameters for consistency, when necessary. Observational errors are not included to avoid clutter. Simulated gradients taken from the literature are shown with lines, \citet[][orange, cross dashed-dotted line for enhanced feedback in MaGGIC g15784 and orange, dashed-dotted line and asterisks for normal feedback in MUGS g15784]{gibson2013} and \citet[][violet, dashed and dotted lines for high and low stellar-mass galaxies $(\mstar > 10^{10} \Msun$ and $\mstar < 10^{10}\Msun$), respectively, in the {\sc fenix} simulation]{tissera2016}.}
\label{gradevol}
\end{figure*}
\begin{table}
\caption{
Redshift evolution of the metallicity gradient, \sohge\, for the star forming gas in the \RECAL\ simulation of the \eagle\ suite. Columns from left to right contain the redshifts, the median values of the  gradient, the  maximum absolute deviations, the bootstrap errors on the medians, the $25^{\rm th}$ and $75^{\rm th}$ the percentiles (all in units of dex~kpc$^{-1}$) and  the number of simulated galaxies at a corresponding z.
}
\begin{tabular}{l|c|c|c|c|c|c}
\hline
z    & $\langle$\sohge$\rangle$ & $\sigma$ & $\sigma_{\rm boot}$ & $25^{\rm th}$ & $75^{\rm th}$ & N \\
\hline                             
   0 & -0.0011  &  0.0373 &  0.0037 &  -0.0197 &   0.0082 &     93\\
0.10 & $\sim 0$ &  0.0415 &  0.0039 &  -0.0133 &   0.0125 &    108\\
0.18 & -0.0004  &  0.0418 &  0.0039 &  -0.0197 &   0.0153 &    103\\
0.36 & -0.0050  &  0.0618 &  0.0063 &  -0.0264 &   0.0165 &    104\\
0.50 & -0.0008  &  0.0540 &  0.0053 &  -0.0273 &   0.0199 &     97\\
0.61 & -0.0055  &  0.0546 &  0.0059 &  -0.0333 &   0.0286 &     84\\
0.73 & $\sim 0$ &  0.0691 &  0.0073 &  -0.0407 &   0.0405 &     86\\
0.86 & $\sim 0$ &  0.0667 &  0.0074 &  -0.0309 &   0.0187 &     85\\
1.00 & -0.0011  &  0.0730 &  0.0091 &  -0.0392 &   0.0298 &     63\\
1.25 &  0.0004  &  0.0768 &  0.0112 &  -0.0519 &   0.0569 &     47\\
1.50 &  0.0008  &  0.0746 &  0.0125 &  -0.0523 &   0.0285 &     37\\
1.73 &  0.0052  &  0.0736 &  0.0114 &  -0.0399 &   0.0584 &     42\\
2.00 & $\sim 0$ &  0.0988 &  0.0177 &  -0.0872 &   0.0465 &     31\\
2.23 &  0.0066  &  0.0592 &  0.0173 &  -0.0281 &   0.0575 &     12\\
2.50 & -0.0056  &  0.1333 &  0.0515 &  -0.0801 &   0.1649 &      7\\
\hline
\end{tabular}
\label{table1}
\end{table}

\begin{figure}
\resizebox{8.cm}{!}{\includegraphics{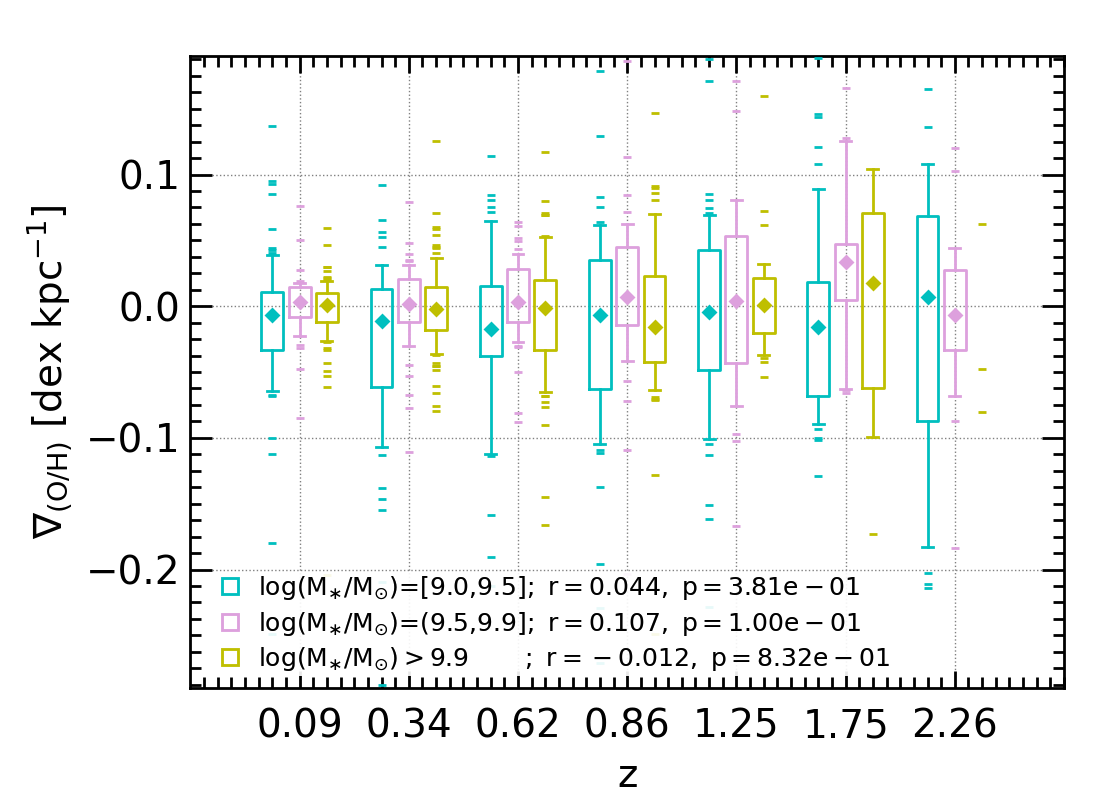}}
\caption{Redshift evolution of the median \sohge\  (diamonds) for galaxies in the three defined stellar mass intervals. The size of the boxes denote the $25^{\rm th}$ and $75^{\rm th}$ percentiles, and the whisker in the boxplot diagrams denote the 10 and 90 percentiles. Data outside of these ranges are plotted individually.}
\label{evolzmasas}
\end{figure}

We plot histograms of \sohge\ normalised to unity in Fig.~\ref{histos}, with colours indicating redshift bin, and using different panels for bins in galaxy stellar mass; key statistics are reported in Table~\ref{table1}. The median gradient is close to zero for all $z$ and $\mstar$. The distribution around the median is fairly symmetric, with outliers up to about $ \pm 0.2~{\rm dex~kpc}^{-1}$.  There is a tendency for the distributions to be skewed towards positive outliers ( i.e., higher metallicity in the outskirts compared to the centre), as indicated by the value of the skewness, $\rm Sk > 1$, reported for each curve.

In contrast, the skewness of  the distributions of the \sohge\ normalised by \Reff  is smaller than one in the interval $z\in[0,0.5]$ (see Appendix~\ref{append:norm}). This indicates a more symmetric distribution,  in agreement with the observational analyses reported by \citet{sanchez2014} and \citet{kewley2010}. At higher $z$, we find that the skewness of the normalised gradient becomes larger than one (see Fig.~\ref{histosnorm}). Hereafter, we will work with \sohge\ rather than normalised slopes, since most of the  current observational samples report metallicity gradients in units of \dexkpc due to the difficulty of measuring accurate galaxies sizes at $z>1$.

We plot the evolution of the median of \sohge\  for all simulated galaxies in Fig.~\ref{gradevol}
as the dark blue line, with the faint blue dashed region including the 25-75$^{\rm th}$ percentile range. The plot compares the \eagle\ results to other simulations (coloured lines) and observations (black symbols). The observational data is included to provide a reference frame  for the simulated data. 
The level of agreement between different observational data sets is quite variable, possibly due to differences in how the galaxies were selected. The mean metallicity gradient and scatter in the data of \cite{wuyts2016} and \cite{curti2019} appear quite similar, with a median close to zero and approximately symmetric scatter around the median, which increases with $z$. The majority of the observed galaxies fall well within the 25-75$^{\rm th}$ percentiles of the simulated galaxies, therefore the simulation agrees well with these two data sets. In contrast, taken at face value, the scatter in \sohge\  is much larger in the data reported by \cite{zaritsky1994} and \citet[][median values for stellar masses are shown]{belfiore2017} at $z\sim0$ as well as that of \cite{carton2018} and \cite{jones2013, jones2015} at high redshift. Clearly, a more in depth comparison of the data requires more a detailed matching of the survey selection criteria.

Figure~\ref{gradevol} also shows the simulated evolution reported by \citet{gibson2013} and \citet{tissera2016}. These authors implemented different subgrid physics prescriptions, in particular varying how supernova (hereafter SN) feedback is accounted for. \citet{gibson2013} analysed the evolution of individual galaxies using two different values of SN energy injected per explosion. The {\sc mugs} simulation uses a lower energy of $4\times 10^{50}$~erg
per SN event, whereas the MaGGIC runs assume $10^{51}$~erg per SN event \citep[see][for more details]{stinson2010,brook2012}. Simulations with the lower injection energy show a stronger trend to have more negative metallicity gradients at higher redshift. The simulation with higher SN energy per event yields shallower gradients, more similar to what is found in \eagle.
The simulation by \citet{tissera2016} adopted the self-regulated SN feedback model of \citet{scan06}, and injects $7 \times 10^{51}$~erg per SN event. They find that the median gradient becomes increasingly negative at higher $z$ for galaxies with $\mstar < 10^{10}~\Msun$; more massive galaxies do not show a clear evolution of the median gradient from $z\sim 2$.\

To quantify the redshift evolution of the metallicity gradients for $z < 2$, a linear regression fit, $y=a+bz$, is applied for \sohge\ as a function of redshift (using a robust least absolute deviation technique). The result shows 
a very weak evolution signal: $a \sim -0.001$~dex~kpc$^{-1}$ and  $b \sim 0.0013$~dex~kpc$^{-1}/\delta z$ (${ r.m.s} = 0.002$ and a Spearman correlation factor of  $\rm r = 0.67, p = 0.0012$).

Additionally, we  calculated the redshift evolution of the metallicity gradients for galaxies in  the three mass intervals defined above.
In Fig.~\ref{evolzmasas} the redshift evolution of the median metallicity
slopes for galaxies in each stellar mass interval is displayed together with the
corresponding $10^{\rm th}$ and $90^{\rm th}$ percentiles.  This analysis was performed by grouping galaxies within a given stellar mass interval, located in two consecutive redshifts,  in order to improve the statistics.
Linear regressions were fitted to each of these relations, finding no clear statistical differences among the three mass groups. The Spearman factors show no significant trend, except for a very weak one for intermediate $\mstar$ galaxies to have shallower gradients at lower $z$.
 They all show a trend to have large diversity at higher redshift. However,  the \sohge\ of massive and intermediate-mass galaxies  exhibit less variety than those of low-mass galaxies for decreasing redshift  as can be appreciated from the $25-75^{\rm th}$ percentiles (or from  the $10-90^{\rm th}$ percentiles). In particular,  low-mass galaxies  are also consistent with a decrease of the dispersion for lower redshift, suggesting an overall flatting of the gradients (indicated by the size of the boxes and whiskers).

\begin{figure}
\resizebox{8cm}{!}{\includegraphics[trim={0 30 0 0}]{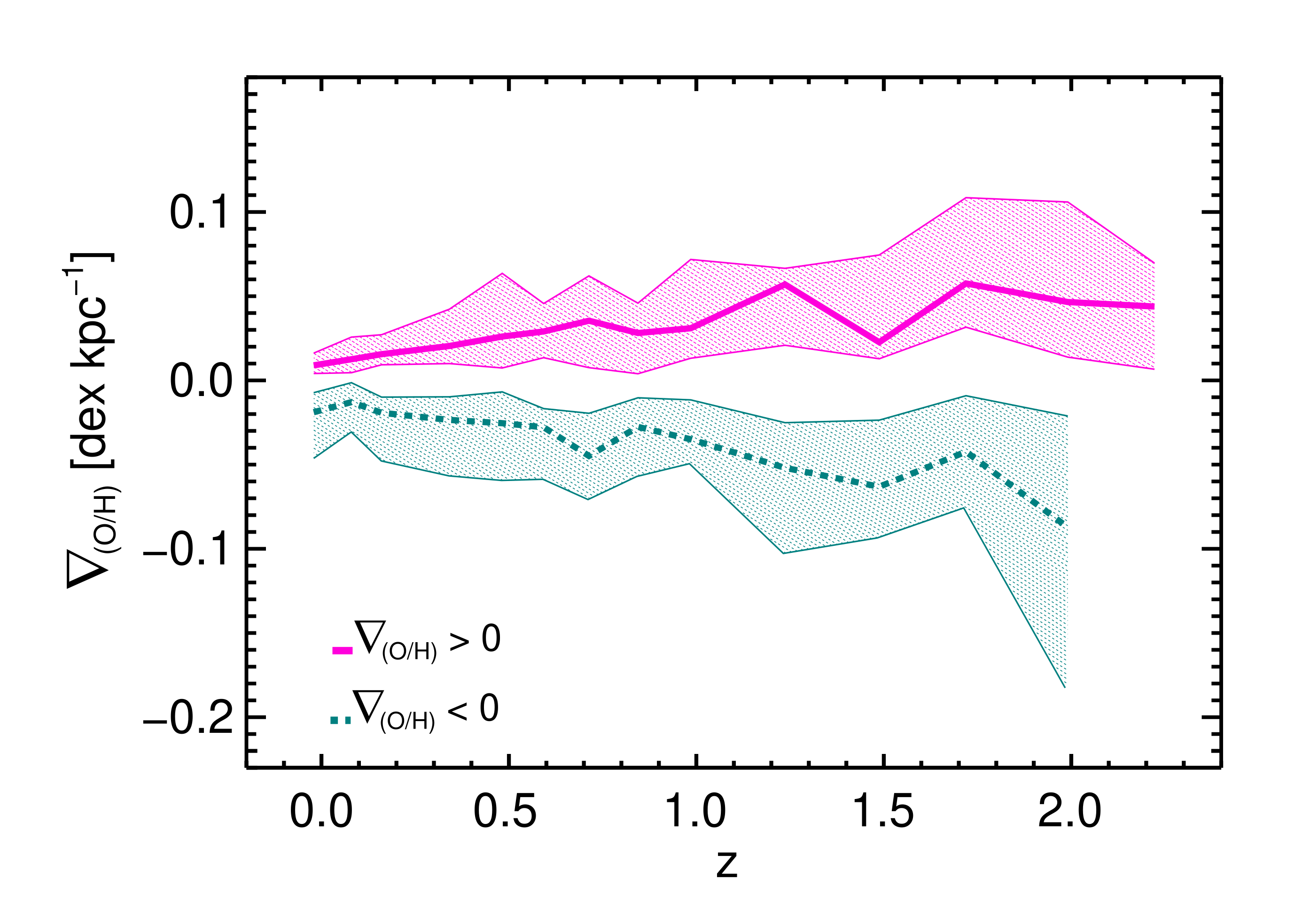}}
\caption{Redshift evolution of the median \sohge\ in galaxies with negative (teal, solid line) and positive (magenta, dotted line). The shaded areas are defined by the 25- $75^{\rm th}$ percentiles.}
\label{diffgradevol}
\end{figure}

To further investigate the evolution of galaxies with positive and negative metallicity gradients, we estimate separately their redshift variation. Figure~\ref{diffgradevol} displays the median gradients (and the first and third quartiles). 
 For positive metallicity gradients, we detect a positive correlation with a slope of  $b \sim 0.019$~dex~kpc$^{-1}/\delta z$ ~$(\rm r=0.83, \ p=0.0004, \ r.m.s = 0.01)$ 
and for the negative metallicity gradients, an anticorrelation with a slope  $b \sim -0.032$~dex~kpc$^{-1}/\delta z$ ~$(\rm r = -0.902, \ p = 5e{-5}, \ r.m.s = 0.01)$.  
The evolution signal detected is slightly stronger for the negative gradients. This level of redshift evolution for negative gradients is in agreement with results of galaxies from the TNG50 simulations reported by \citet{hemler2020}. These authors find  negative metallicity gradients, which yield a redshift evolution of $-0.02 $~dex~kpc$^{-1}/\delta z$, {\rm which is consistent with observations for $z < 0.5$, but they do not detect a preference for flat gradients for  $z > 1$}.

We also calculated the fractions of galaxies with positive metallicity slopes as a function of redshift.
 If we use $\sohge = 0$ to segregate them,  similar percentages of about 45-52 per cent are found at all  redshifts analysed. When we adopt the median gradients as reference and estimate the fraction of positive  gradients above the medians by $+3\sigma_{\rm boot}$, the percentages of positive gradients represent about $\sim 20-25$ per cent.

Hence, the weak positive redshift evolution is not driven by an increase in the fraction of galaxies with positive gradients but   by the fact that the variety of metallicity gradients becomes larger, with some of them being steeper, for higher redshift (at a given galaxy size as we will show later).
It remains to be determined if the fractions of positive gradients detected in \eagle\ are in agreement with observations. This information will provide clues to improve the subgrid physics in galaxy formation models.

\subsection{The stellar mass dependence of the metallicity gradients} \label{sec:grad_mass}

\begin{figure}
\resizebox{8cm}{!}{\includegraphics[trim={0 10 0 0}]{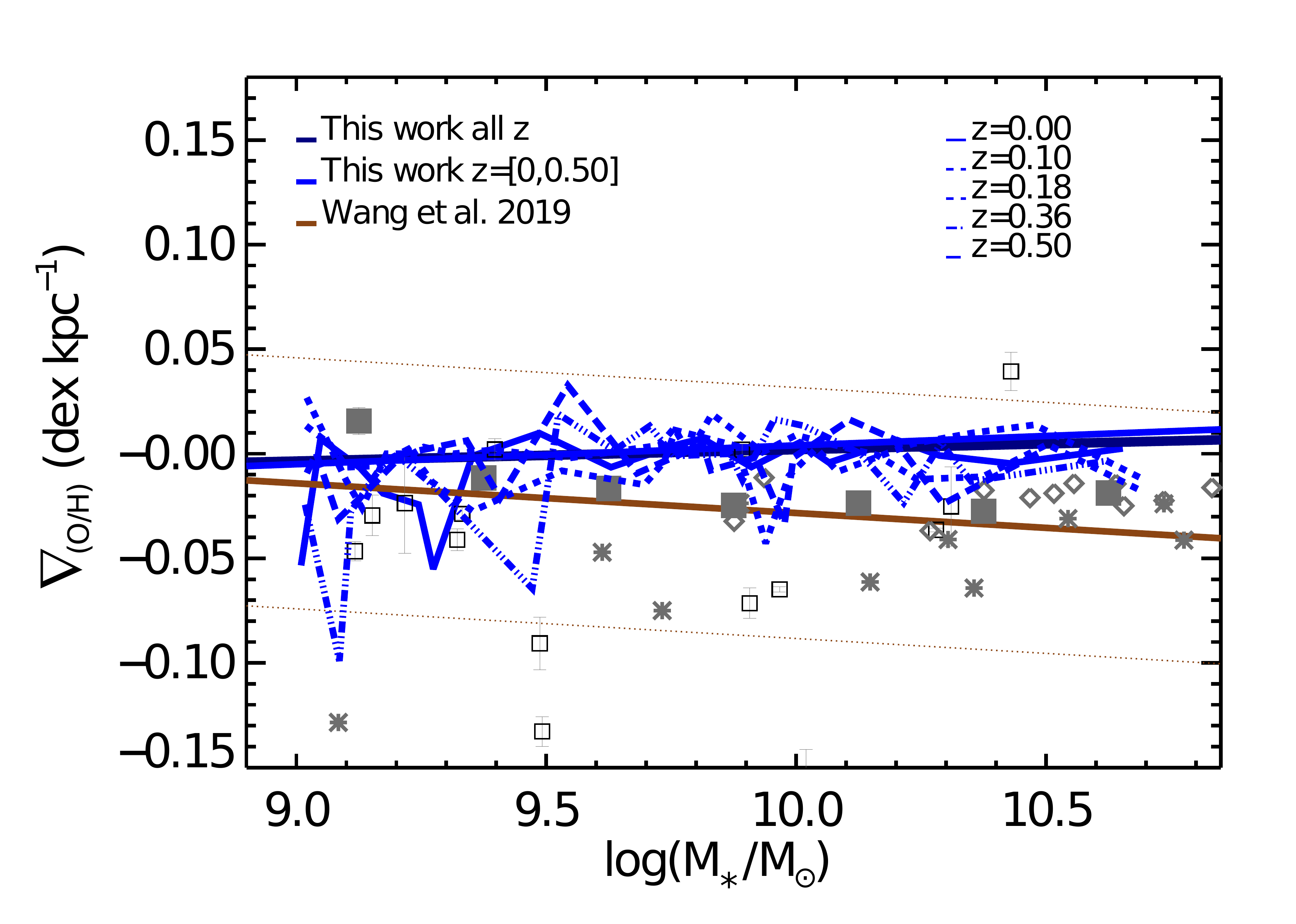}}\\
\resizebox{8cm}{!}{\includegraphics[trim={0 10 0 0}]{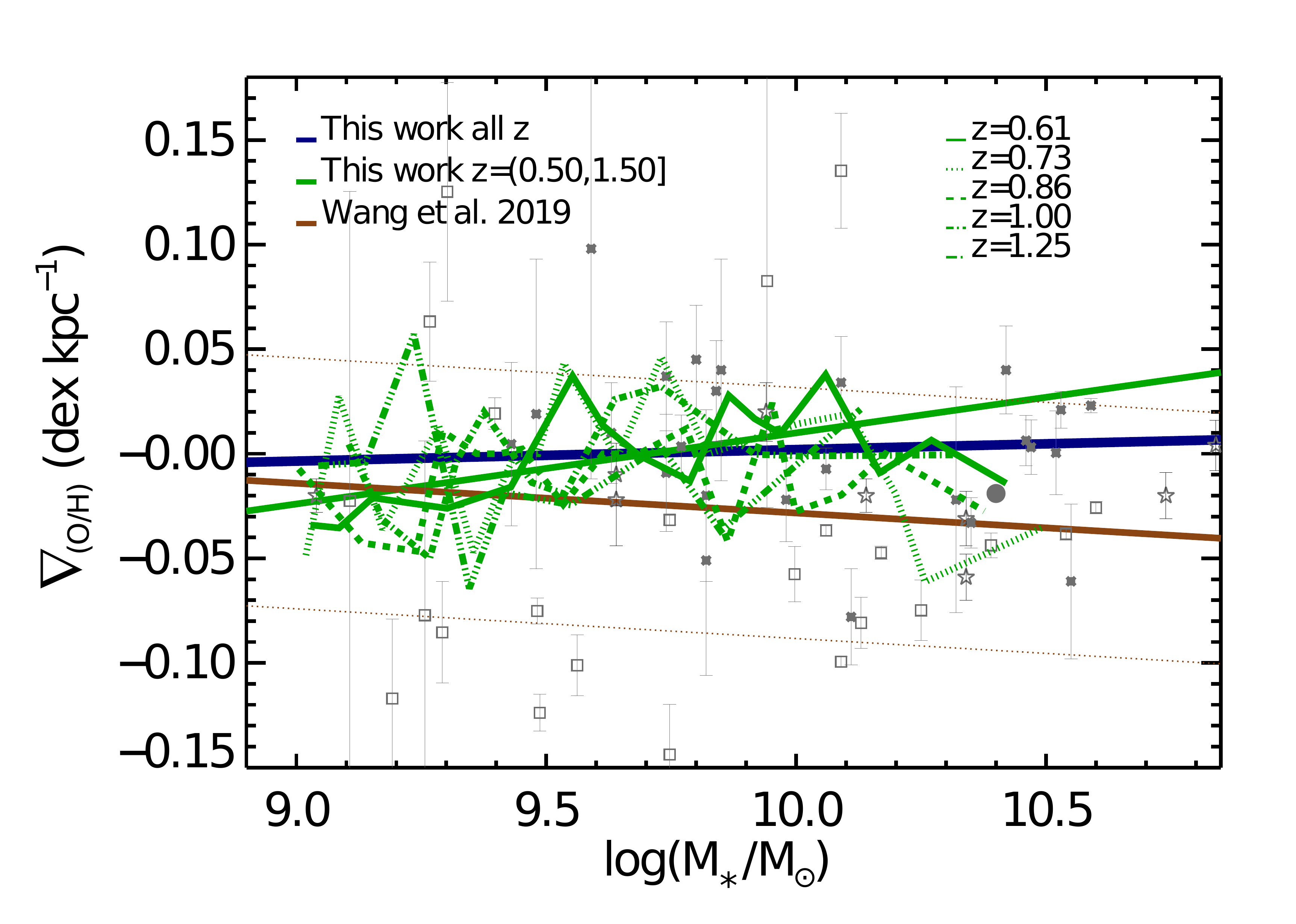}}\\
\resizebox{8cm}{!}{\includegraphics[trim={0 50 0 0}]{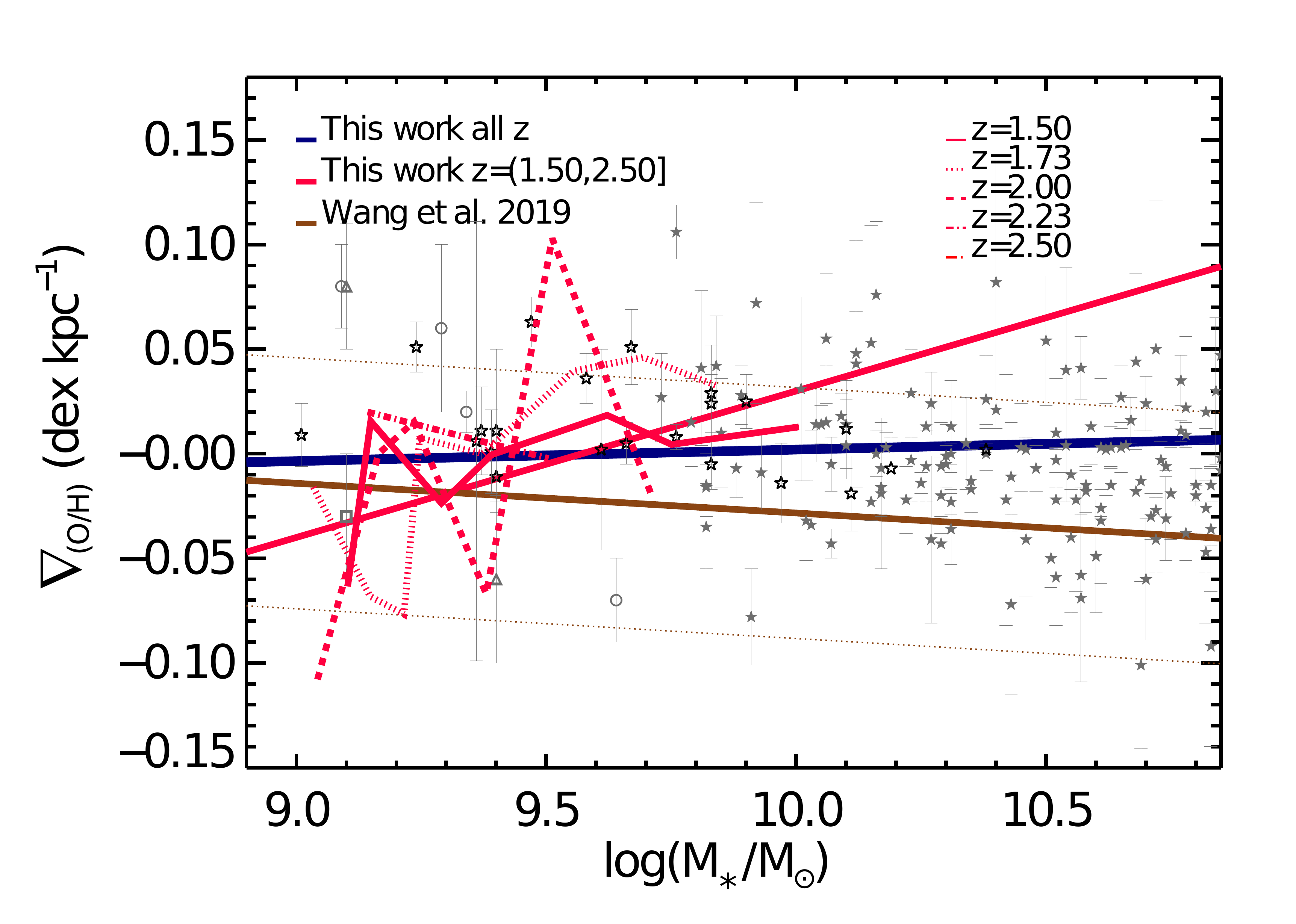}}\\
\caption{Median \sohge\  
as a function of $\mstar$ of the selected \eagle\ galaxies in the three defined redshift intervals as indicated by the  labels. 
The linear regression to both the gradients within a given redshift range (blue, green and red lines) and the overall collected gradients for $z \leq 2$ (dark blue, thick solid lines) are included.
The median relations are also depicted at each analysed redshift to visualise the level of variation among them. For comparison, the observational relation reported by \citet[][brown, solid lines]{wang2019b} and other observational data have also been  included (see Fig.~\ref{gradevol} for symbol code).
}
\label{gradmass}
\end{figure}

In this section, we explore the relation between the \sohge\ and the galaxy stellar mass as a function of redshift.
 First, we built a median \sohge-$\mstar$ relation by collecting the  simulated galaxies with $\rm M_{star} \leq 10^{10}~\Msun$ within each of the  three defined redshift intervals and median \sohge-$\mstar$ relations at each analysed redshift. An upper $\mstar$ limit is imposed in order to sample the same mass range in all $z$, considering that, at  high redshift, there are few galaxies with $\mstar  > 10^{10}~\Msun$ in our sample as shown in Table~\ref{table1}.

Then, we performed linear regressions on \sohge-$\mstar$ for each $z$ interval, using the formula: $\sohge\rm = \alpha + \beta\ log(\mstar/\Msun)$.
In the case of overall  \sohge-$\mstar$ (i.e. using all galaxies within $z = [0,2]$), the linear regression yields $\alpha = -0.053$ \dexkpc and $\beta = 0.006$ \dexkpc
$\rm (r.m.s = 0.007)$. In this case, a correlation signal is detected with a Spearman correlation factor of $r = 0.25$ ($p = 0.06$), indicating a weak trend for more massive galaxies to have shallower metallicity gradients.

The \sohge-$\mstar$ relations are shown in Fig.~\ref{gradmass}. 
 In each panel, the median values for each analysed redshift (as indicated by the inset labels) as well as the median relations for galaxies within each  redshift interval are included. The linear regression fitted to the latter are also depicted. 
For comparison, the linear regression for all galaxies within $z = [0,2]$ (dark blue, solid lines) 
and the  linear relation reported by \citet{wang2019b} (brown, solid lines) of observed galaxies  within $z = [1.2,2.3]$, are also shown. As can be seen from this figure and Table~\ref{table2}, EAGLE galaxies show $\beta \geq 0 $ for galaxies with $\mstar = [10^9,10^{10}]~\Msun$.

The weakening of  the \sohge-$\mstar$ relation for decreasing redshift  (Table~\ref{table2}) indicates that, at a fixed stellar mass, galaxies have shallower metallicity gradients at lower redshift and, that 
the change in slope is larger for lower mass galaxies as also shown in Fig.~\ref{evolzmasas} and~\ref{gradmass}.
This result is in agreement with the redshift dependence of \sohge\ for galaxies with different stellar masses shown in Fig.\ref{evolzmasas}.  We note that the signals are nevertheless  very weak. The positive dependence of \sohge\  on stellar mass for galaxies with $\mstar > 10^9 \rm M\sun$ is in agreement with the estimations from  semi-analytical galaxy formation models by \citet{yates2020}. However,  it seems at odds with some observational results   \citep[e.g.][]{belfiore2017,poet2018} which show an anti-correlation  for $\mstar  \leq 10^{9.5} \rm M\sun$ but in agreement with others \citet{bresolin2019}.

\begin{table}
\caption{ Linear regression parameters for the stellar mass dependence of the metallicity gradients (\sohge $=\rm \alpha + \beta\ log(\mstar /10^{9.5}\Msun)$) in the three defined redshift intervals. Columns from left to right contain $\alpha$, $\beta$, maximum absolute deviation ($\sigma$), the {\em r.m.s} (all in dex~kpc$^{-1}$) and the Spearman coefficients (r, p). The linear regressions have been applied within the mass interval} $\mstar = [10^9,10^{10}]~\Msun$.
\centering
\begin{tabular}{c|c|c|c|c|c}
\hline
z interval & $\alpha$ & $\beta$ & $\sigma$& {\em r.m.s} & r \\ 
$[0.00, 0.50]$ & -0.003 & 0.01 & 0.01 &0.02 & 0.20 (p= 0.050)\\
$(0.50, 1.50]$ & -0.006 & 0.03 & 0.01 & 0.02 & 0.19 (p= 0.100)\\
$(1.50, 2.50]$ & -0.009 & 0.07 & 0.01 & 0.04 & 0.47 (p= 0.002)\\
\hline
\end{tabular}
\label{table2}
\end{table}

\subsection{Galaxy size dependence of the metallicity gradients} \label{sec:grad_size}

As shown by T19,  the $z=0$  \sohge\  in EAGLE exhibits a clear correlation with $R_{\rm eff}$  and the morphology of galaxies (i.e., more disc-dominated galaxies tend to have extended discs with shallower metallicity gradients; see figure 4 in T19). 
Recently, \citet{bresolin2019} reported that both low-and-high surface brightness galaxies defined a relation between metallicity gradients and the galaxy size at $z=0$.
We note that these galaxies show negative \sohge. This observed relation is weaker than that proposed by \citet{prantzos2000}, based on a pure inside-out formation model. This discrepancy between models and observations could be explained by considering that galaxies formed in a more complex scenario  where other physical mechanisms, such as mergers, gas inflows and  radial mixing, among others, take place. If this is the case, as also speculated by \citet{bresolin2019}, then the relation for our simulated galaxies,  formed within a cosmological context where such processes are taken into account,  should reproduce  the observed relation.  However, this would be expected for those with negative \sohge, but it might not be true for those with positive gradients as we will discuss below.

\begin{figure}
\resizebox{8cm}{!}{\includegraphics[trim={0 10 0 0}]{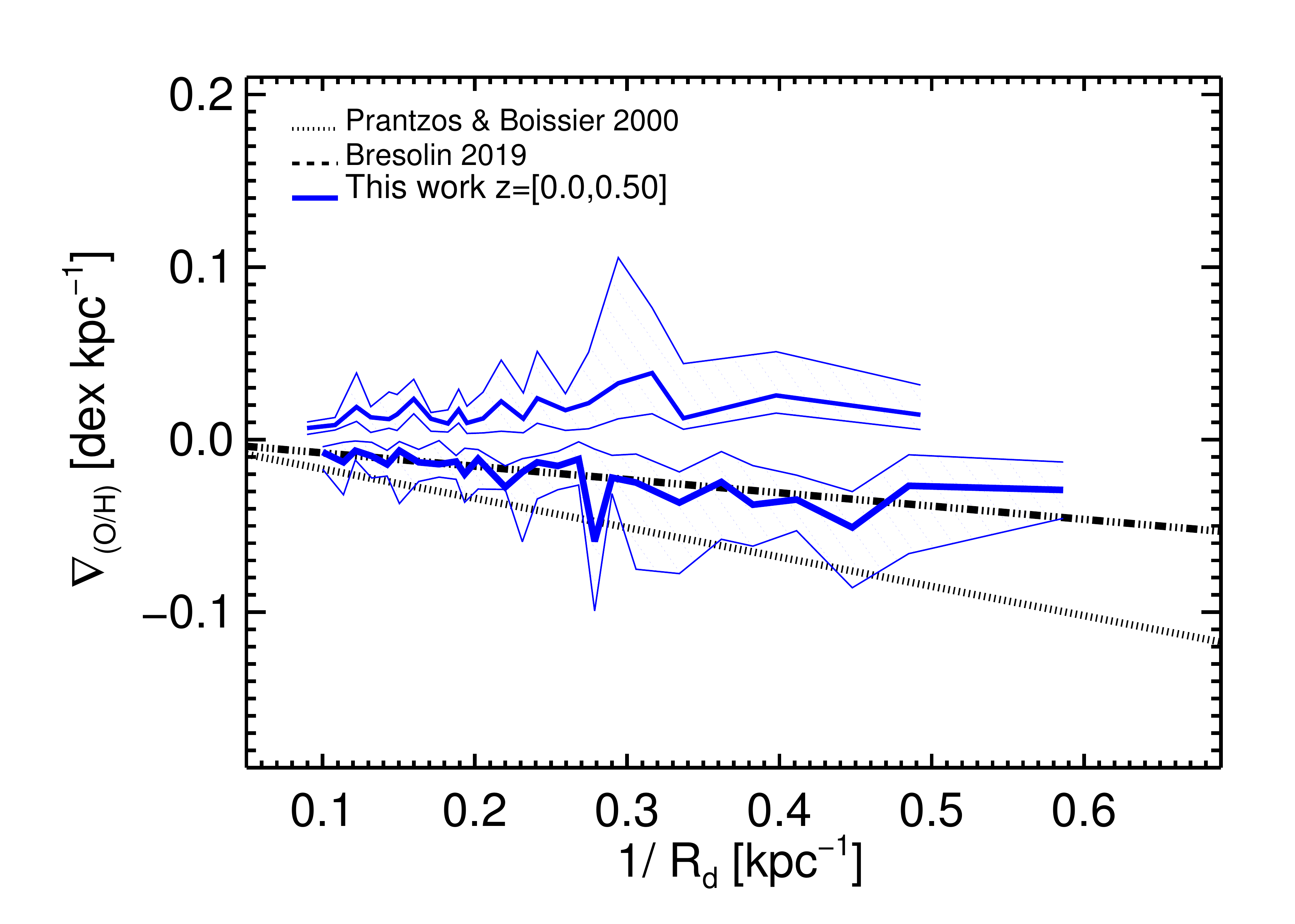}}
\resizebox{8cm}{!}{\includegraphics[trim={0 10 0 0}]{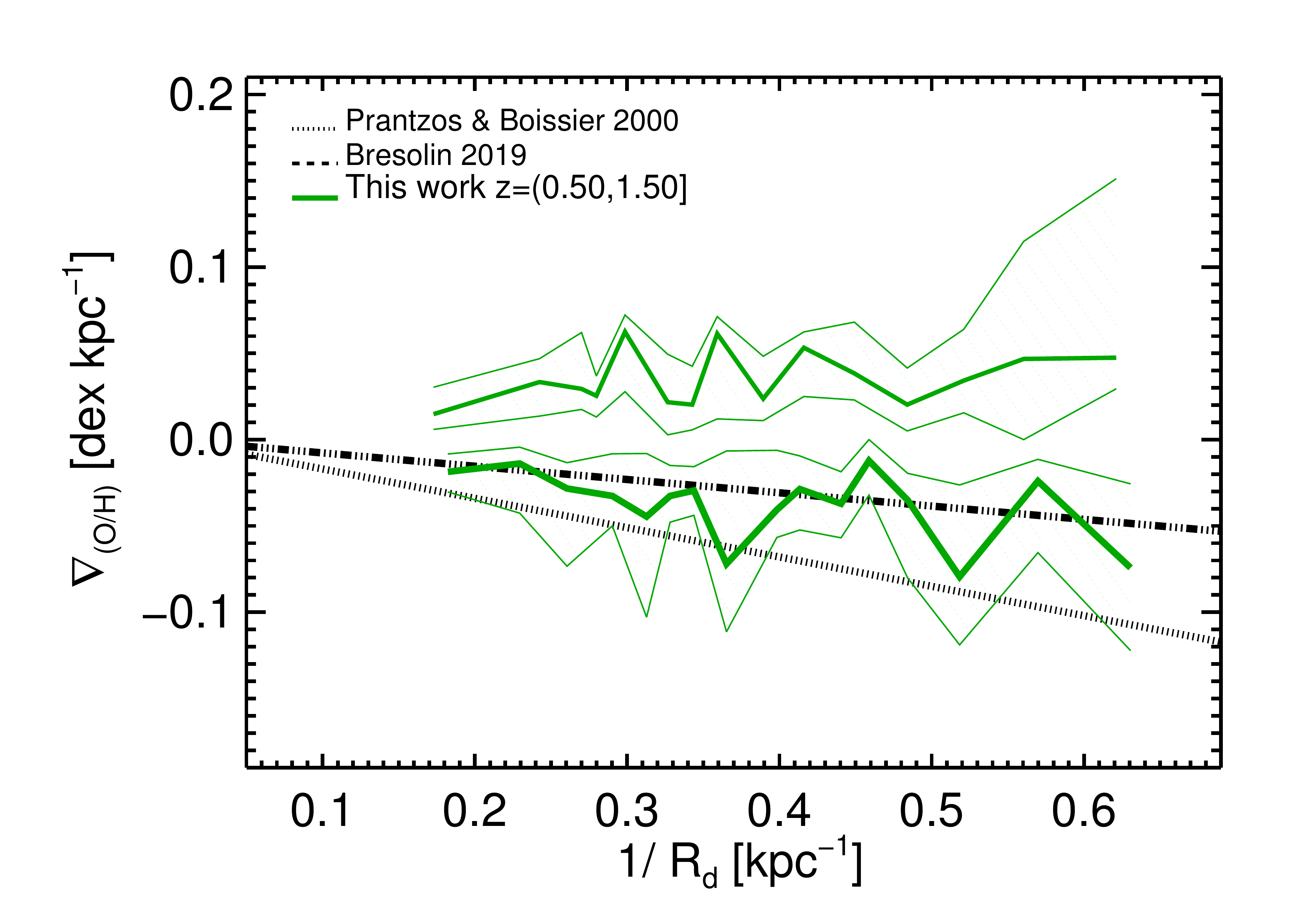}}
\resizebox{8cm}{!}{\includegraphics[trim={0 20 0 0}]{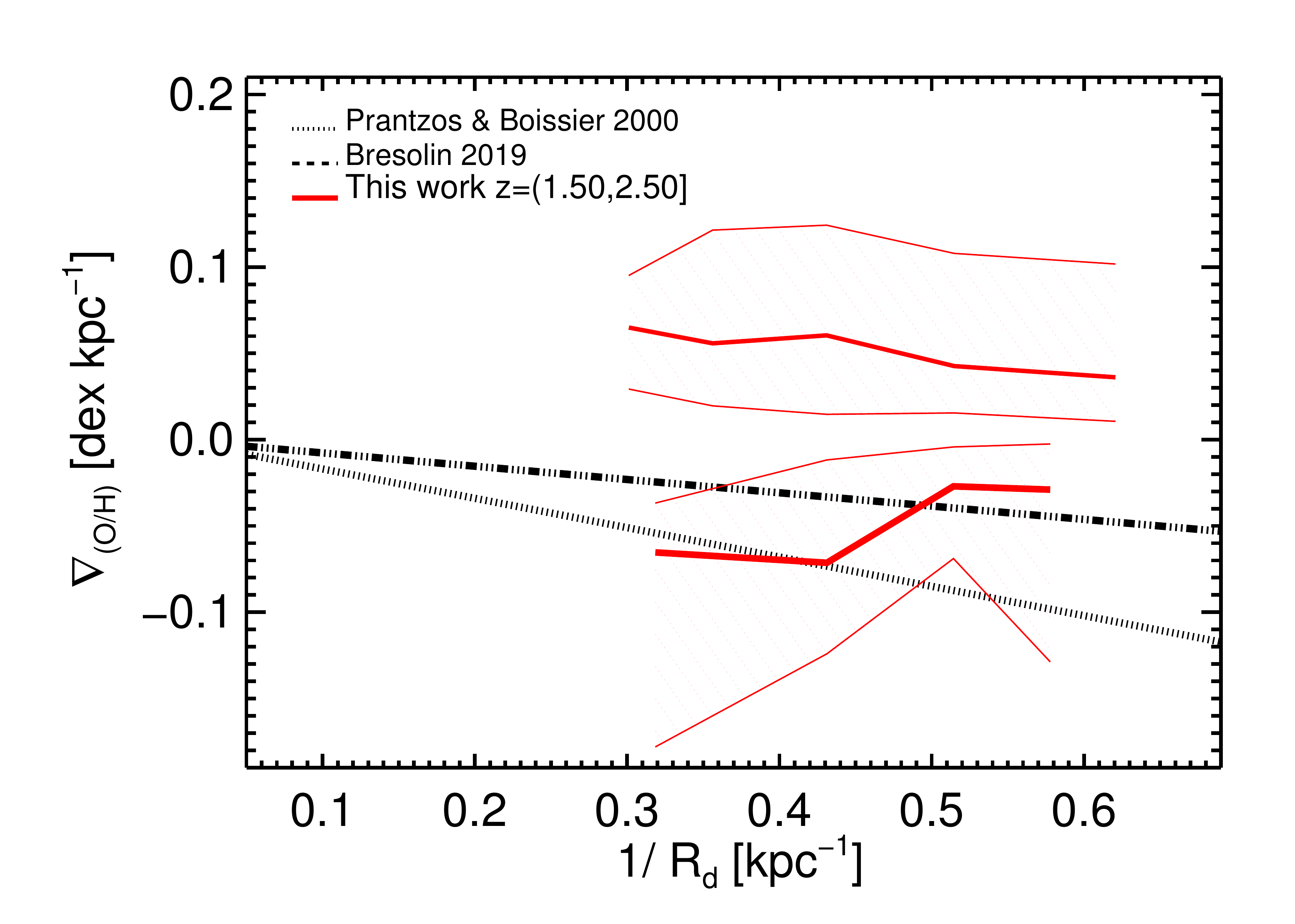}}
\caption{Median \sohge\  as a function of the inverse of the stellar disc scale-length for the EAGLE galaxies with positive (upper distributions) and negative (lower distributions) metallicity slopes in the three defined redshift intervals (as indicated in each  panel). The shaded areas  delineate  the $25^{\rm th}$ and $75^{\rm th}$ percentiles. 
For comparison, the observed relation reported by \citet[][black dashed-dotted lines]{bresolin2019} and the analytical estimation by \citet[][black dotted lines]{prantzos2000} for galaxies with $\sohge<0$ are also included.}
\label{gradvsreff}
\end{figure}

To  compare with the observational results of \citet{bresolin2019}, we estimated $R_{\rm d}$ by assuming $R_{\rm d} = 1.86\ R_{\rm eff}$ (see Section 2.2).
Collecting the simulated gradients within each redshift interval, we calculated  the global median relations for galaxies with positive and negative \sohge, separately.
As can be seen from Fig.~\ref{gradvsreff}, there is a clear anti-correlation between \sohge\  and the inverse of disc size for galaxies with negative metallicity slopes at all  redshifts analysed. 
In fact, the simulated trends are in excellent agreement with \citet[][black, dashed-dotted line]{bresolin2019} at $z \sim 0$.
Conversely, positive \sohge\  determine a flat/positive correlation with $R^{-1}_{\rm d}$. 

 We applied linear regressions of the form \sohge $= \alpha + \beta R^{-1}_{\rm d}$ to the medians relations determined by  galaxies with positive and negative gradients within each redshift interval.
For $z \in (0.50, 1.25]$, the negative dependence for $\sohge < 0$ is slightly weaker than for lower redshifts and has  larger dispersion (see Table~\ref{table3}).
For both subsamples in the highest redshift interval, the linear regressions  yield opposite tendencies compared to those estimated for  lower redshifts. However, considering the significant decrease in  the number of members in the highest redshift intervals (see Table~\ref{table1}), we do not consider these fittings robust.

\begin{table}
\caption{Fitting parameters for negative and positive metallicity gradients as a function of the inverse galaxy scale: redshift intervals, the slope of the linear regressions ($\beta$ in dex), the maximum absolute deviation ($\sigma$ in \dexkpc) and the Spearman coefficients.}
\centering
\begin{tabular}{l|c|c|c|c|l}
\hline
 &  z {\rm interval} & $\beta$ & $\sigma$ & r \\
                         &$[0.00, 0.50]$ & -0.086 & 0.006 & -0.81~(p$=7e{-7}$) \\ 
$\nabla_{\rm (O/H)} < 0$ &$(0.50, 1.50]$ & -0.049 & 0.012 & -0.39~(p$=0.130$)     \\
                         &$(1.50, 2.50]$ & ~0.140   & 0.008 & ~0.35~(p$=0.400$)     \\
\hline
                         &$[0.00, 0.50]$ & ~0.054 & 0.006 & ~0.54~(p$=0.009$)  \\
$\nabla_{\rm (O/H)} > 0$ &$(0.50, 1.50]$ & ~0.060 & 0.010 & ~0.35~(p$=0.190$)    \\
                         &$(1.50, 2.50]$  & -0.090 & 0.002 &-0.90~(p$=0.040$)  \\
\hline
\end{tabular}
\label{table3}
\end{table}

The \sohge-\Reff\ relation for galaxies with negative $\sohge< 0$ results in  global agreement  with the observational trend reported by \citet{bresolin2019} who find  $\beta \sim -0.077$ dex at $z \sim 0$. The simulated relation does not evolve  significantly with redshift up to $z \sim 1.5 $ as can be seen from Fig.~\ref{gradvsreff}.
Indeed, as galaxies become smaller in size and stellar mass for increasing redshift\citep{furlong2015}, they move along approximately the same relation.

The good agreement of the simulated relation  for galaxies with negative \sohge\  with observations and, hence, the discrepancy found  with the analytical predictions of \citet{prantzos2000} are likely to be due to the fact that the EAGLE galaxies formed in a cosmological scenario as a result of a balance between inflows, outflows, mergers and interactions as expected \citep[e.g.][]{sommer1999, pedrosa2015, teklu2015}.  
Our results for negative \sohge\  agree  with those obtained by \citet{hemler2020} for TNG50 simulations and Tronrud et al. (in preparation) for the Auriga simulations.

We also note that an increase in the dispersion of the metallicity gradients in \dexkpc as a function of redshift is expected as galaxies become smaller \citep{carton2018, bresolin2019}. However, as can be seen from Fig.~\ref{gradvsreff}, at a given galaxy size, the variety of \sohge\ increases with increasing redshift. Hence, the large diversity of \sohge\ might also   reflect physical variations in the metallicity distribution in the gas-phase at different cosmic times.

The dissimilar trends found in galaxies with positive and negative gradients at a given galaxy scale, suggest that, at the time gradients are calculated, galaxies with positive \sohge\  might be affected by  physical mechanisms that could redistribute the gas and the metals in the ISM in a different manner than those with negative \sohge. As a result,  the link between size  and \sohge\  expected in an inside-out scenario  disappears, as we will discuss in the following sections.

\subsection{sSFR dependence of the metallicity gradients}

\begin{figure}
\resizebox{8.cm}{!}{\includegraphics[trim={0 0 0 0}]{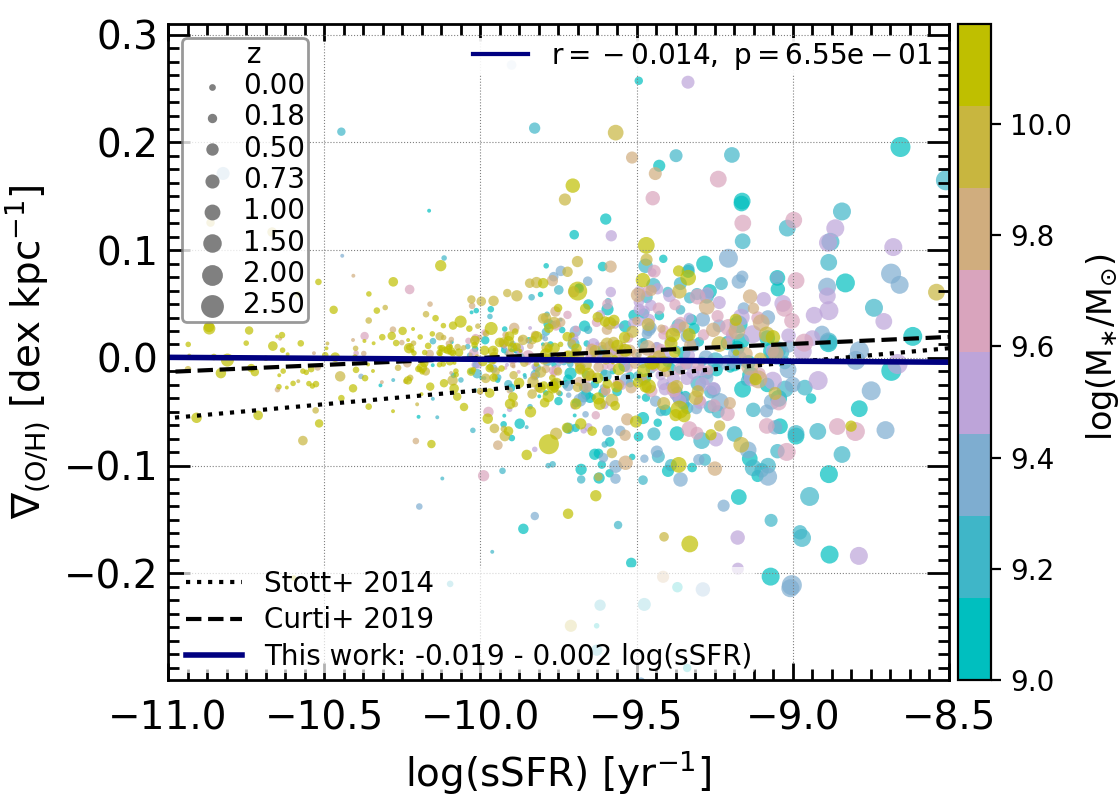}}
\resizebox{8.cm}{!}{\includegraphics[trim={0 0 0 0}]{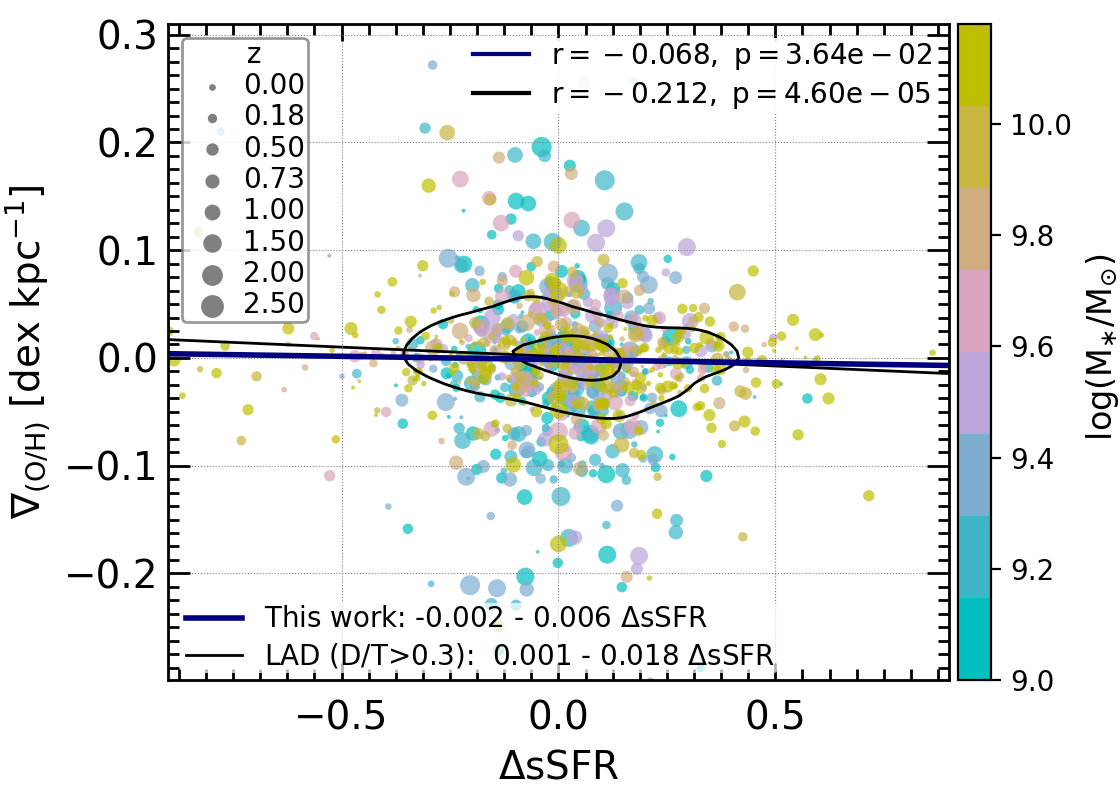}}
\caption{
Upper panel: Distribution of \sohge\ as a function of sSFR for galaxies selected with  $z \in[0, 2.5]$. For comparison, the linear regressions fitted to the simulated distributions (blue solid line), and the observed relations reported by \citet[][black dashed lines]{stott2014} and \citet[][black dotted lines]{curti2019} are depicted. The colours represent  $\mstar$. Lower panel: Distribution of \sohge\  as a function of $\rm \Delta sSFR$. Here, the colours represent the galaxy morphology as indicated by D/T. The black line depicts the linear fit for galaxies with $\rm D/T > 0.3$ and the black contours enclose the $25^{\rm th}$ and $75^{\rm th}$ per cent of the distribution.
}
\label{ssfr_grad}
\end{figure}

In this section we  explore the relation between \sohge\  and sSFR \citep{stott2014, wuyts2016, curti2020}.
In the upper panel of Fig.~\ref{ssfr_grad}, we show \sohge\  as a function of sSFR, for all analysed redshifts  (depicted by the size of the symbols). The colours display the values of $\mstar$ of the simulated galaxies. 
Higher sSFR are systematically found for increasing redshift in agreement with the evolution of the sSFR-M$_{\rm star}$ relation.
The median relations and linear regression fittings of the form 
 \sohge $\rm = \alpha + \beta\ log(sSFR/10^{-9.0} yr^{-1})$ are also displayed.
 As can be seen, we find no clear dependency ($\rm \beta \sim -0.002\ dex\ kpc^{-1},\ r=-0.014$).

The lower panel of Fig.~\ref{ssfr_grad} shows the metallicity gradients as a function of $\Delta{\rm sSFR} = \rm log(sSFR/\langle sSFR(M_{\ast},z)\rangle)$, defined as the distance from the median main sequences determined by our galaxy sample at each  redshift. As can be seen, there is no clear trend as  quantified by the Spearman coefficients are  $\rm r=-0.068\ (p=3.64 \times 10^{-2})$. This is also the case when a similar analysis is performed for  galaxies with different $\mstar$ with negative and positive  \sohge, separately.

We interpret the  lack of correlation in our EAGLE sample as  mainly due to the large diversity of galaxies with different \sohge\  at a given sSFR or $\Delta{\rm sSFR}$.
\citet{sillero2017} analysed how mergers affected the sSFR and \grad, concluding that only sudden gas inflows with time-scales smaller than 0.5~Gyr triggered by  mergers could simultaneously produce an  increase in  the sSFR and a flattening of  the metallicity gradient, as suggested by \cite{stott2014}. Unsurprisingly, the gas fraction of the merging galaxies affects both the sSFR and the value of the gradient, as do the orbital parameters of the merger \citep{torrey2012}. In contrast, an isolated galaxy that does not undergo major mergers usually evolves more smoothly with redshift, converting  steadily the gas reservoir into stars. This   contributes to enhance the chemical content in the ISM, modifying the metallicity gradient  gradually towards a shallower one.   Hence,  when galaxies with  different  formation histories are combined together, the correlation is likely to be suppressed or weakened due to the large variety of evolutionary tracks \citep{tissera2016}, even if it exists  for individual galaxies  \citep{ma2017grad,yates2021}.

Other factors can also contribute to reinforce the lack of dependence on sSFR in this simulation. On the one hand,  the EAGLE simulations produce a lower log sSFR by $0.2-0.5$~dex than expected from observations at $z > 1$ \citep{furlong2015}, and hence, at high redshift strong sSFR might be missing from our sample.  On the other hand, observed galaxy samples might be biased towards high mass systems with high star formation activity, principally at $z>3$, which contributes to setting the observed relation (a redshift range which is beyond our analysis).

However, more disc-dominated galaxies with  $\rm D/T > 0.3$ show a weak anti-correlation between \sohge\  and $\rm \Delta sSFR$ with $\rm \beta \sim -0.018~dex~kpc (r=-0.212,\ p=4.6 \times 10^{-5})$ as can be seen in the lower panel of Fig.~\ref{ssfr_grad}. This finding indicates that disc-dominated galaxies tend to have  more negative \sohge\ if they are actively forming stars while those that are more quenched tend to have shallower ones. These results are in agreement with those reported by T19, who found that disc galaxies with negative \sohge\ have, on average, more recent star formation activity (within the last ${\sim}1.5$ Gyr) than those with positive slopes.

\section{The impact of mergers on the metallicity gradients} \label{sec:mergers}

\begin{figure*}
\resizebox{16cm}{!}{\includegraphics{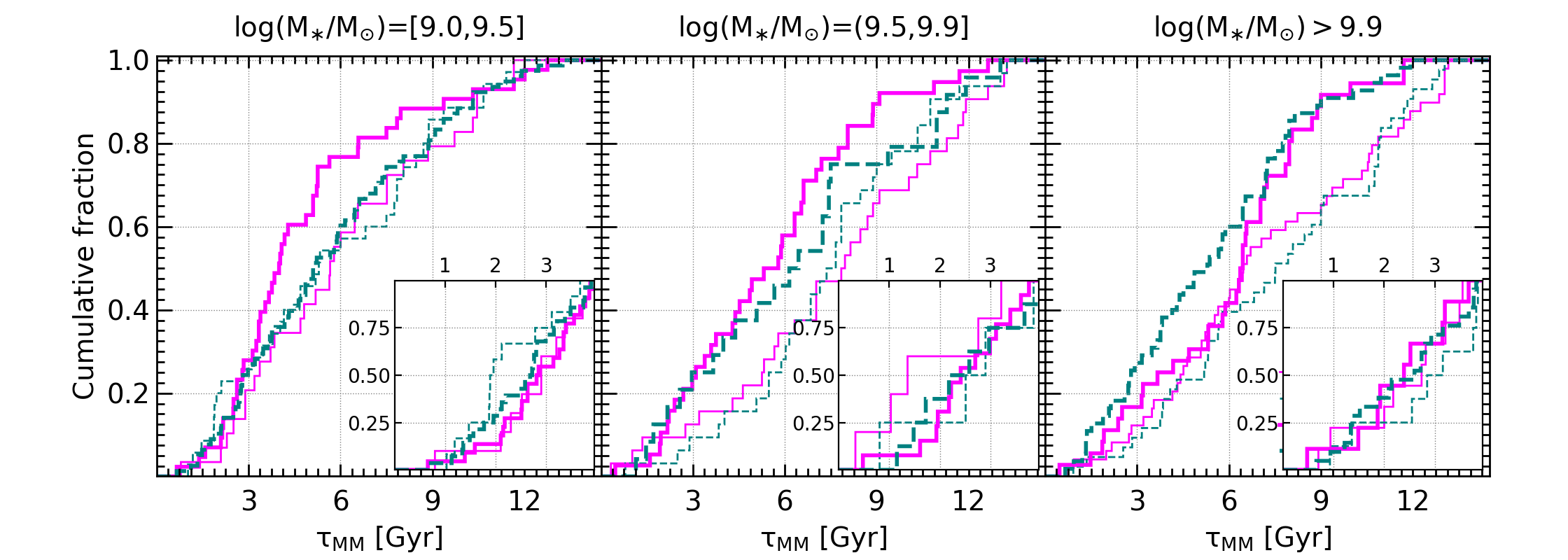}}
\resizebox{16cm}{!}{\includegraphics[clip,trim={0 0 0 62}]{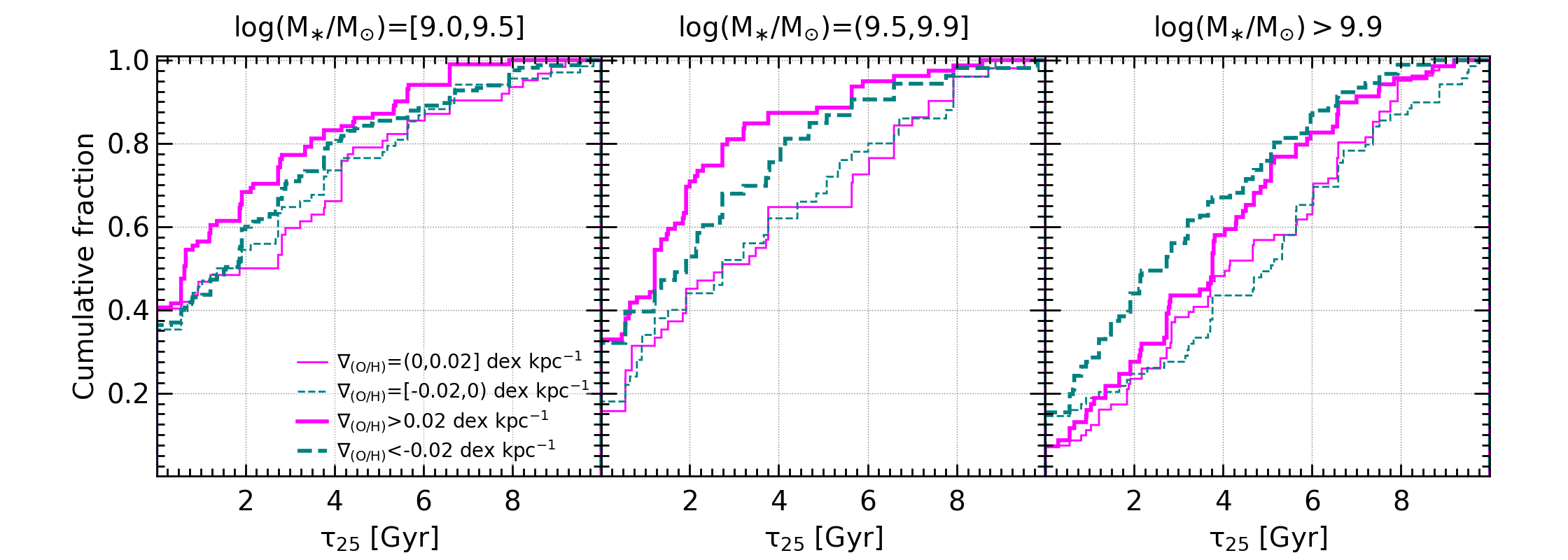}}
\caption{ Upper panel: Cumulative galaxy number fractions as a function of the time since the last major merger event, \taumm, \ 
for galaxies in the three defined stellar mass intervals. Colours denote the \sohge\  subsample to which they belong as indicated in the legends. The insets show the  cumulative distributions for systems with \taumm$ < 4$ Gyr.
Lower panel: Cumulative galaxy  number fractions as a function of time since the last major starburst \taustar\ displayed as above.
Galaxies with weak gradients tend to have more recent quiescent star formation histories (i.e. longer \taustar) and fewer recent major mergers (i.e. longer \taumm) compared to  those with strong negative or positive \sohge.}
\label{mergertime}
\end{figure*}

In this section, we  analyse the properties of galaxies with positive and negative \sohge\ as a function of their recent merger histories by exploring their merger trees. The star formation activity, morphology and gas-richness of the progenitors of our galaxies  are estimated as a function of time.

From the merger trees,  the most recent merger events are selected. Due to the sparse number of available outputs, we adopt as the merger time the last one where  two galaxies are identified as separated entities by the {\sc SUBFIND} algorithm. This time provides a reference date for a merger since the actual collision takes place  between two consecutive snapshots (separated by at most 1 Gyr).
With this information, we estimate the time, \taumm, elapsed since the merger  and the time when the \sohge\ is computed. 

Mergers are classified as minor or major according to the stellar mass ratio, 
$R_{\rm merger} = \rm M_{\ast,\ accreted}/M_{\ast,\ prog}$, between an accreted  satellite and a galaxy. We define 
minor mergers as those within  $R_{\rm merger} = [0.1,0.25]$ and major ones as $R_{\rm merger} \ge 0.25$. In this paper, we will only consider major mergers since they are found to produce a larger impact than  minor ones. 
The \sohge\ and all the analysed properties are calculated for the progenitor galaxies at \taumm.

For  each galaxy, we also estimated the time, \taustar, elapsed since the last major increase in stellar mass by at least 25 per cent.
In this case, we are agnostic about the 
physical processes that produced such increase  in  stellar mass, which could be  minor or major mergers, interactions or gas inflows.

To highlight the trends, we grouped galaxies according to their metallicity gradients in four subsamples: 
strong positive and negative metallicity gradients are classified as those with $\sohge > 0.02$~\dexkpc and $\sohge < -0.02$~\dexkpc, respectively, 
and weak positive and negative metallicity gradients as those with $\sohge \in [0,0.02]$~\dexkpc and $\sohge \in [-0.02,0]$~\dexkpc, respectively. 
The threshold values of $\sohge = 0.02$ (-0.02)~\dexkpc are chosen to have approximately comparable number of members in each subsample.

\subsection{The impact on metallicity profiles}

 In the upper panels of Fig.~\ref{mergertime},  we display the cumulative number fractions of galaxies in the four defined galaxy subsamples as a function of \taumm. 
As can be seen, there is a trend for galaxies with strong \sohge\ to have  undergone more recent merger events. Galaxies with $\rm \mstar \leq 10^{9.9} \Msun$ favour strong positive gradients associated with more recent active merger histories (i.e. lower \taumm), while more massive galaxies show an excess of strong negative \sohge\ for similar \taumm. In the inset,  we concentrate on the latest 4 Gyr to highlight the effects of more recent mergers. Intermediate  and high mass galaxies show an excess of positive gradients for mergers taken place within the last $\sim2$ Gyr. 

These trends agree with those obtained for the  cumulative fractions of galaxies as a function of \taustar, shown in the lower panels of Fig.~\ref{mergertime}. As can be seen, galaxies with strong positive gradients tend to have  smaller \taustar, followed by those with strong negative gradients. For massive galaxies, this is also valid  but the relations for positive and negative strong gradients switched position, with strong negative metallicity gradients preferring galaxies with more recent stellar mass growth.

In summary, for $z \leq 2$, our findings show that galaxies with weak (positive or negative) \sohge\ tend to have  more quiescent star formation and merger histories compared to those with strong \sohge, regardless of their stellar mass.

\subsection{The impact on galaxy properties}

\begin{figure*}
\resizebox{16cm}{!}{\includegraphics[trim={0 20 0 0}]{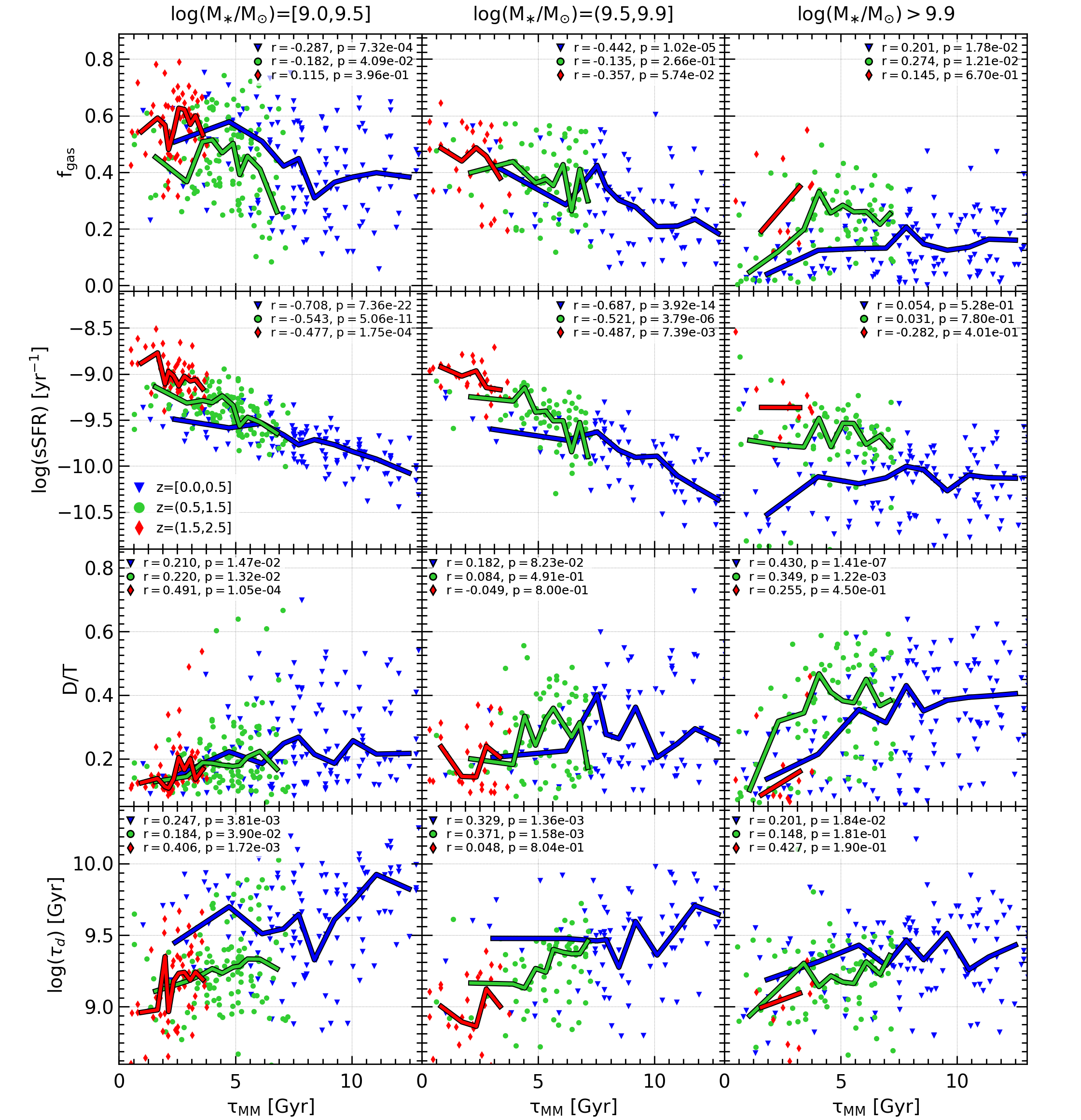}}
\caption{Distributions of \fgas, sSFR,  D/T and $\tau_{\rm d}$ as a function of the time of the last major merger, \taumm\, for galaxies with low, intermediate and high stellar masses (left, middle, and right panels). In each panel, the distributions (symbols) and the medians (solid lines) of properties  are coloured according to the three defined redshift intervals (indicated in the inset labels). The Spearman correlation factors are also included.
}
\label{tau_prop}
\end{figure*}

We also explored the dependence of \fgas, sSFR, D/T and depletion time, 
$\rm \tau_{d} = M_{\rm gas}/SFR$, on \taumm \ of  a galaxy after a merger event as shown in Fig.~\ref{tau_prop}.  A similar analysis was performed by using \taustar \ instead as displayed in Fig.~\ref{tau25prop}. For this purpose,  galaxies  are grouped by   $\mstar$ and redshift. For each subsample, the median trends are computed  by adopting a variable binning in $\rm \tau_{MM}$, so that each bin includes at least five galaxies. The number of bins is allowed to vary within the range $\rm n_{bins}=[2,10]$. The median relations for a given subsample are only calculated and shown if $\rm n_{bins} > 2$.  We note that this is a statistical analysis performed by using
all galaxies identified with the adopted criteria and that all identified mergers are displayed regardless of their \taumm \ in order to clearly visualise the effects related with recent mergers without imposing a strict time threshold.

In the upper panels of Fig.~\ref{tau_prop}, \fgas \ is shown as a function of \taumm .
As expected, galaxies are more gas-rich at higher redshift independently of  $\mstar$.
For low and intermediate mass galaxies,  there is  a clear anticorrelation  between \fgas\ and \taumm, indicating   systematically higher \fgas\ for galaxies with recent mergers. This suggests that the accreted satellites bring in gas and/or that  the main galaxy is  gas-rich itself. This is even more plausible for higher redshifts, where galaxies are, on average, more gas-rich as can be also seen from this figure. Alternatively,  mergers could disturb the ISM so that transformation of gas into stars is less efficient \citep{davis2015}. However,  the sSFR  of galaxies with  low and intermediate $\mstar$ also shows a strong dependence on \taumm\ so that galaxies with recent mergers tend to have higher sSFR. This is shown in the second row of Fig.~\ref{tau_prop} (the trend persists if  ${\rm \Delta sSFR}$ is considered instead of sSFR).  

Massive galaxies with $\mstar \geq 10^{9.9}\rm M_{\sun}$ have lower \fgas\ at all $z$, as expected. Contrary to lower mass galaxies, their gas reservoirs decrease even more  during {\bf recent} merger as indicated by the correlation found with  \taumm \ (upper, right panel of Fig.~\ref{tau_prop}).  This suggests that during the mergers the gas reservoir is converted efficiently into stars and/or ejected via SN/AGN feedback, quenching these galaxies. As can be seen from the second row of Fig.~\ref{tau_prop}, they show a flat sSFR median trend, with only a few galaxies reaching high star formation activity. Our results agree overall with those of \citet{hani2020} who reported similar trends for the star formation activity in the TNG300-1 simulations, albeit for  $\mstar \geq 10^{11.5}\rm M_{\sun}$.

Galaxy interactions and mergers are also known to disturb galaxy morphology, destabilising the discs, and contributing to the formation of dispersion-dominated systems. Hence, we expect D/T ratios to anti-correlate with \taumm \ (and \taustar) so that galaxies with more recent mergers become less disc-dominated. Indeed, the  third panel of Fig.~\ref{tau_prop} displays the relation between these parameters where an anticorrelation between the median D/T and \taumm \  can be appreciated (see also Fig.~\ref{tau25prop}). The impact of mergers on galaxy morphology is found at all analysed redshift and stellar masses  for small \taumm \ (< 2-3 Gyr).

The last row of Fig.~\ref{tau_prop} shows $\rm \tau_{\rm d}$ as a function of \taumm. As can be seen, there is a global redshift evolution of $\rm \tau_{\rm d}$ for all masses towards longer depletion times for lower redshift as they become less gas-rich and have lower star formation activity. However, for all $z$ and $\mstar$, $\rm \tau_{\rm d}$ decreases for lower \taumm, suggesting that  major mergers play a significant role in increasing the efficiency of the transformation of gas into stars, regardless of the  \fgas \ (see upper panels).


\subsection{Central metallicities}

 We inspect the level of enrichment of the central regions for the simulated galaxies with respect to the median values of systems with similar stellar masses, located at the same redshift. This estimation provides an indication of  any  dilution of the central abundances in the case of a decrease with respect to the median. Conversely, the enhancement of central abundances with respect to the median  suggests  higher metal retention fractions or/and sustained  star formation activity. Of course, we expect galaxies with positive gradients {\bf to} have lower central metallicity than those with negative ones, on average. With this analysis, we aim to quantify this trend.

For this purpose, we estimated the oxygen abundance at $\rm 0.5 R_{eff}$ with respect to the median oxygen abundance of galaxies at a given stellar mass and redshift, 
$\rm \Delta (OH)_c = (O/H)_c - \langle(O/H)_c (\mstar,z)\rangle$. Hence, for each redshift, the medians of $\rm (O/H)_c$ for each $\mstar$ interval are calculated. Then, for each galaxy at that redshift, the $\rm \Delta (OH)_c$ {\rm is} estimated with respect to the corresponding median. In this way, the dependence on the global evolution of the metallicity as a function of stellar mass and redshift {\bf is} eliminated. Hence,  $\rm \Delta (OH)_c$ reflects the variations of the central level of enrichment due to other physical mechanisms such as mergers.

\begin{figure}
\resizebox{8cm}{!}{\includegraphics[trim={0 10 0 0}]{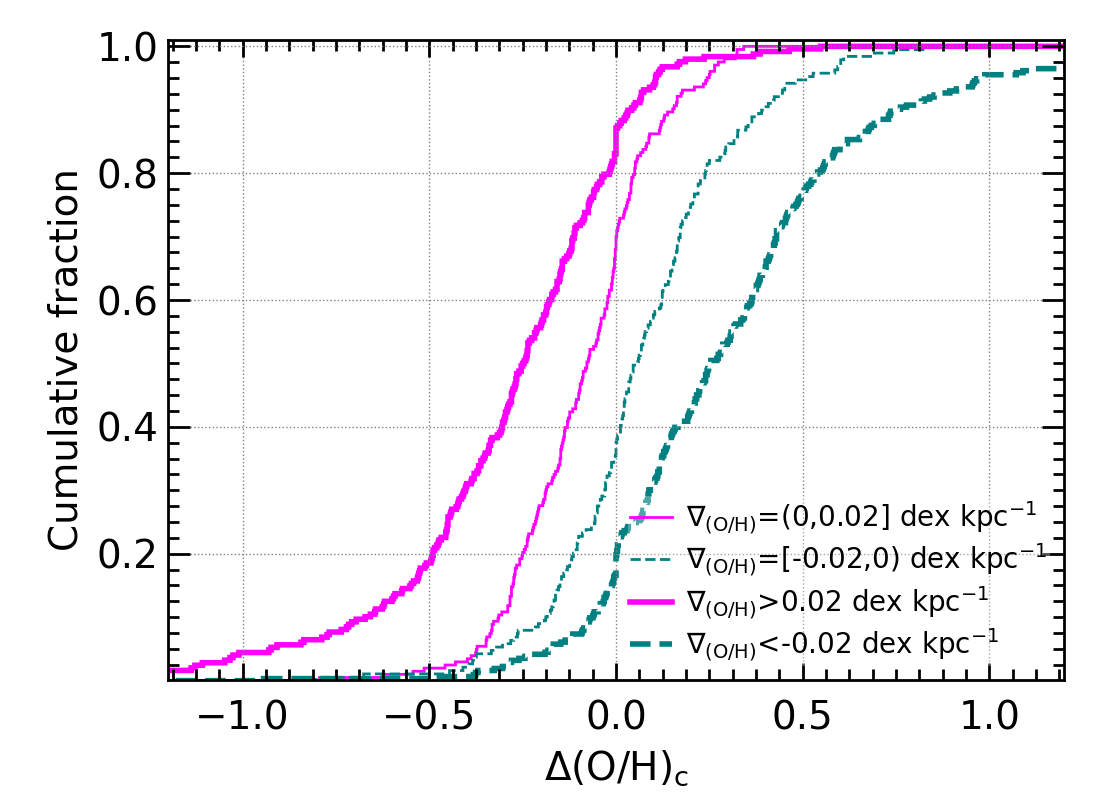}}
\caption{Cumulative distribution of  $\Delta{\rm (OH)_c}$ for the four defined metallicity gradients (given in \dexkpc) groups. A systematic trend for galaxies with stronger positive gradients to have lower central metallicity than those with negative slopes is found as expected if low metallicity gas inflows dilute the central abundances and/or feedback mechanisms took place, ejecting enriched material. Regardless of mass and redshift, galaxies with negative gradients have  higher retention factors and/or sustained star formation activity. Galaxies with positive and negative weak gradients  still show significant enhancement or dilution, respectively. }
\label{centraloh}
\end{figure}

Figure ~\ref{centraloh}  shows  the cumulative distributions of  $\rm \Delta (OH)_c$ for the four defined gradient subsamples. Each distribution has been normalised to the corresponding number of members. As can be seen, there is a systematic trend for strong and weak positive gradients to have lower central metallicity, whereas  galaxies with negative \sohge\  tend to have more enhanced central abundances. Although we detect larger $|\rm \Delta (OH)_c|$ for galaxies with stronger gradients, those with weak ones still show significant metal dilution or enhancement. The significant $\rm \Delta (OH)_c$ detected could be related to the subgrid physics implemented in the simulations and hence, provides another constraint to improve it.

\subsection{Comments on the positive metallicity gradients}

 A significant fraction (20-25\%) of  positive metallicity gradients, which remains approximately constant for $z \leq 2$, are detected in the \eagle\ galaxies.  These systems can be explained by the action of mergers which can trigger low-metallicity inflows and strong starbursts capable of inducing metal-loaded outflows via SN/AGN feedback. 

Positive metallicity gradients could be also related to positive age profiles since the new stars formed in the inner region would contribute to flattening or inverting the age slopes, depending on the relative contributions of coeval stellar populations. Previous works on dispersion-dominated systems in the \eagle\ simulations showed a fraction of rejuvenated systems with positive age gradients \citep{rosito2019b, rosito2019}, which seems to be in excess compared to observations \citep{lagos2020}. On the other hand, \citet{varela2021} analysed the age profiles and mass surface density of disc-dominated system in an {\sc EAGLE} simulation, and reported the existence of a significant fraction of disc-dominated galaxies with on-going central star-formation activity and positive age profiles.

Hence, the determination of the frequency of these features in nature can render further constraints on the subgrid physics implemented in simulations (see T19 for a discussion of the impact of SN feedback on the metallicity gradients).
This information is also key  to understand the possible long-standing impact of  galaxy interactions and mergers on the regulation of star formation and metallicity distribution in galaxies. 
In fact, the parameters of the subgrid physics in \RECAL were recalibrated to improve the description of  the galaxy stellar mass function at $z=0$ \citep{schaye2015}. This enhanced feedback could imprint a more intense variation in the chemical enrichment of the central regions compared to weaker feedback as shown in T19. 


We also note that gas-rich galaxies  tend to have  clumpier ISM, a phenomenon more frequent at high redshift. As a consequence, the metallicity distribution in the ISM can be more heterogeneous and hence, the azimuthal-averaged estimations of the metallicity profiles could hide important information \citep[see][for example]{sharma2019,solar2020}. This could be overcome through the use of spatially  resolved relations.

\section{Evolution of the metallicity gradients of individual galaxies } \label{sec:individual}

\begin{figure*}
\resizebox{17.5cm}{!}{\includegraphics[trim={0 0 0 0}]{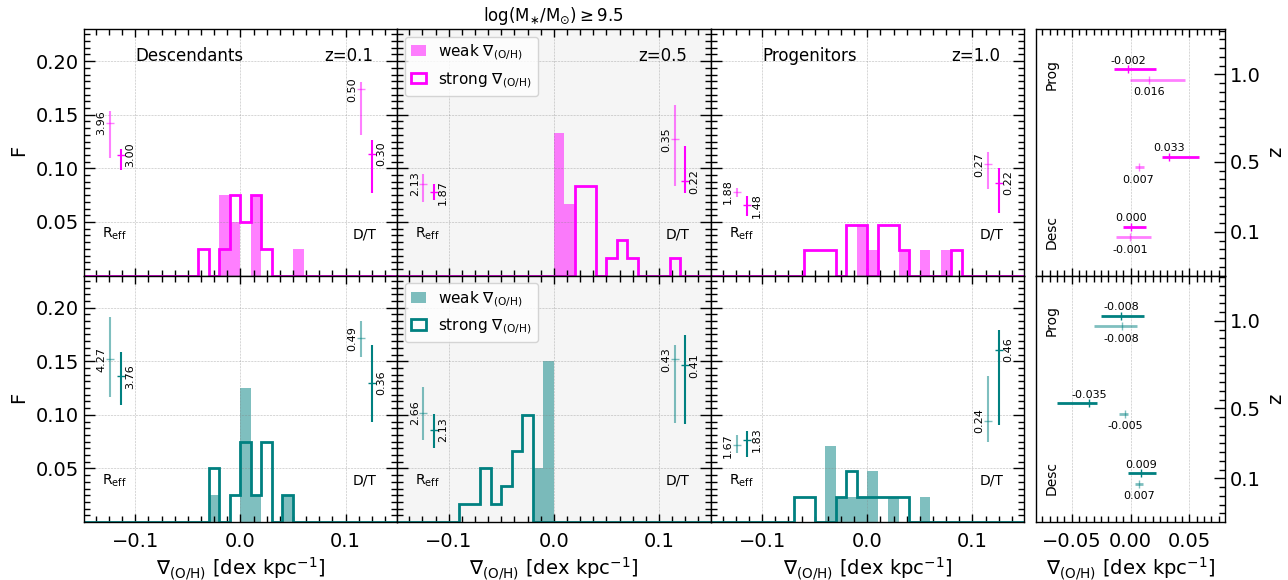}}
\caption{Histograms of \sohge\  for galaxies with $\mstar \geq 10^{9.5} \rm \Msun$ selected at $z=0.5$. Top panel: galaxies with strong (magenta, solid lines) or weak (magenta, shaded areas) positive \sohge\  at $z=0.5$ (second column) and their descendants (first column) and progenitors (third column). Bottom panel: as above but for galaxies with strong (teal, solid lines) and weak (teal, shaded areas) negative \sohge. In each panel the median values of \Reff\ and D/T have been included (the size of the bars denote the $25$-$75^{\rm th}$ percentiles). The fourth column depicts the median values of \sohge\ and the $25^{\rm th}$ and $75^{\rm th}$ percentiles for each histogram (as error bars).
}
\label{histogradselected}
\end{figure*}

Our findings suggest that as 
the merger rates and \fgas \ increase with higher redshift,  the frequency of strong positive and negative metallicity gradients also rises  as discussed in  the previous section. As a consequence, the variety of metallicity gradients  becomes  larger for higher redshift (Fig.~\ref{gradevol}). 
This trend  is in agreement with previous works on the evolution of the  metallicity gradients in individual galaxies, which reported  large scatter and rapid variation of the gas-phase metallicity gradients for higher redshift \citep[see also][]{ma2017,yates2021}. 


Here, our goal is to trace the evolution of \sohge\ for individual galaxies  along their merger trees, while keeping a statistical approach. For this purpose,  we selected galaxies with strong/weak and  positive/negative gradients at $z=0.5$ and search  for their progenitors at $z=1.0$  and their descendants at $z=0.1$,  using the merger trees in the \eagle\  database described by \citet{mcAlpine2016}. 
Hence, all galaxies selected at $z=0.5$ have $\mstar > 10^{9.5} \rm M_{\sun}$ and a progenitor at $z=1.0$ and/or a descendent at $z=0.1$ with measured \sohge. We work in this redshift range  because we found the largest number of galaxies with both descendent and progenitors that satisfy our selection criteria and hence, have measured metallicity gradients (see Section \ref{sec:sims})

Figure~\ref{histogradselected} shows the distributions of positive and negative \sohge\  (upper and lower panels, respectively). In each panel, the strong (solid lines) and weak (shaded areas) types are displayed separately. The fourth column of Fig.~\ref{histogradselected} presents the medians of \sohge\ for each type (the length of the bar is determined by the $25^{\rm th}$ and $75^{\rm th}$ percentiles).
As can be seen, galaxies with strong positive or negative \sohge\ evolve from progenitors with a larger diversity of slopes, although both subsamples show a tail towards positive and negative values, respectively.
After passing through  evolutionary stages where the \sohge\ become  steeper, their descendants converge to shallower metallicity gradients at lower $z$. 

Galaxies with weak positive and negative \sohge\ have progenitors with a diversity of  gradients, but their distributions are skewed towards  strong positive or negative \sohge, respectively, as quantified by the $25-75^{\rm th}$ percentiles.  They converge to metallicity profiles with shallow slopes already by $z\sim0.5$ and  retain the weaker slopes down to $z\sim0$. 
The rate of change of the medians \sohge\ is larger for strong types, $\rm \sim 0.01 ~dex~kpc^{-1} Gyr^{-1}$, than for weaker ones $\rm \sim 0.0003 ~dex~kpc^{-1} Gyr^{-1}$. 

As the metallicity gradients evolve towards $z\sim 0$ so do the sizes and morphologies of the galaxies.
The medians of the \Reff\ and D/T of each subsample (the length of the bars is determined by the $\rm 25^{\rm th}-75^{\rm th}$ percentiles) are also displayed in Fig.~\ref{histogradselected}. Globally, galaxy sizes grow with time from similar small median \Reff\ at $z = 1$, to more extended systems at lower redshifts.
Galaxies with weak positive gradients show an increase in \Reff\ by a factor of $\sim 2$, with those with negative do so by a factor of  $\sim 2.6$. Galaxies with strong negative and positive \sohge\  also grew by a factor of two approximately.  However,  galaxies with  negative gradients are systematically more extended.  

In Fig.~\ref{histogradselected}, the evolution of morphological indicator, D/T, is also included. As can be seen,
galaxies with strong, negative gradients tend to be more disc-dominated
than those with strong, positive slopes. The latter are more dispersion-dominated at all analysed time. Galaxies with weak negative and positive evolve systematically towards more disc-dominated systems, with the former increasing their D/T by a factor of two approximately. 

Galaxies with negative \sohge\  are more disc-dominated at all analysed times compared to those with positive ones. This trend is consistent with the well-defined galaxy-size dependence of \sohge\  for these galaxies, discussed in Section~\ref{sec:grad_size}.  As shown in the previous subsection, they also favoured enriched metallicity central regions compared to those positive gradients as expected.

\begin{figure}
\resizebox{8cm}{!}{\includegraphics[trim={0 0 0 0}]{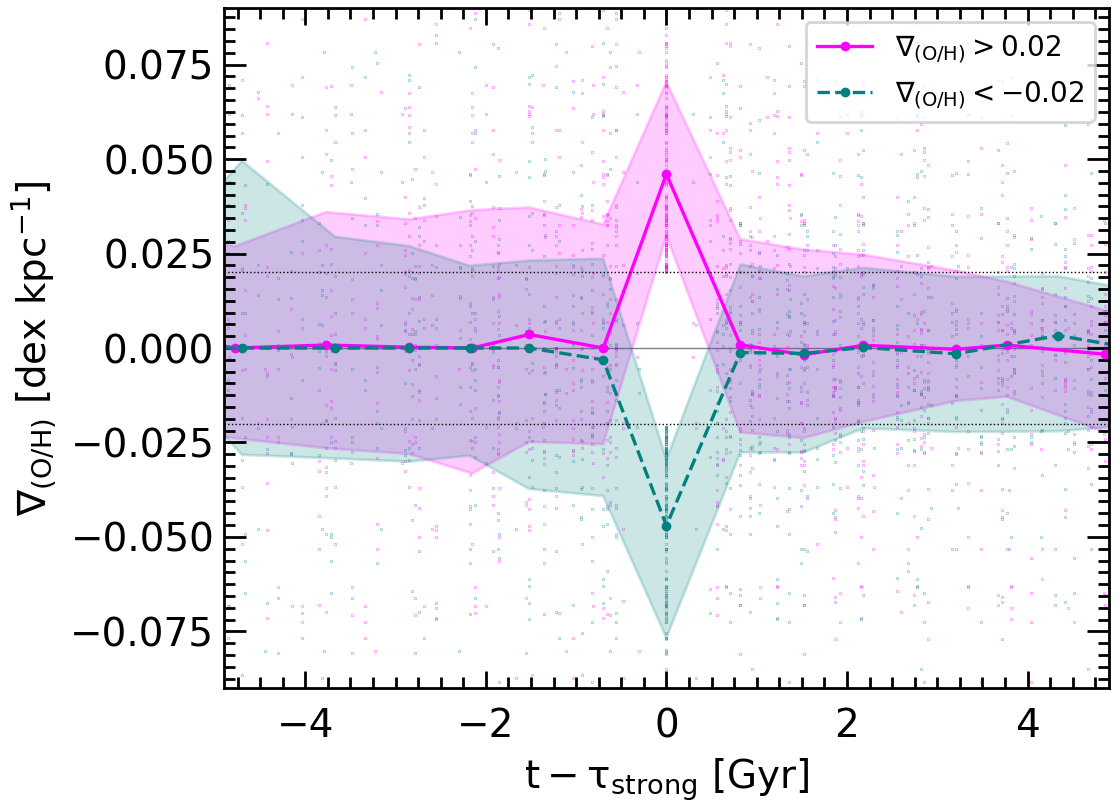}}
\caption{Evolution of the median \sohge\  for galaxies with $\mstar \geq 10^{9.5} $ as a function of $\tau -\tau_{\rm strong}$, the elapsed time since gradients become strong negative  (teal, dashed lines) or strong positive (magenta, solid lines). 
The shaded regions   delineated  the $25-75^{\rm th}$ percentiles. 
}
\label{gradreset}
\end{figure}

In order to assess the typical timescale of variation of the metallicity gradients from strong to weak types, first we identified the redshfit at which  galaxies and  their progenitors, along their merger trees, had strong negative or positive \sohge. Then the time is reset to zero at that redshift ($\tau - \tau_{\rm strong} = 0 $). In this way, we can stack the evolution history of the metallicity gradients synchronising them 
at the times where their slopes are strongly enhanced. In Fig~\ref{gradreset} the median evolution of the \sohge\ as a function of $\tau - \tau_{\rm strong}$ is displayed for strong negative  (teal, dashed lines) and strong positive (magenta, solid lines) gradients. 
We find that strong metallicity gradients are a transitory stage in the evolution of galaxies with a median timescale of $2 $~Gyr. The 25th  and the 75th percentiles are at $1.4$ Gyr and $\sim 3$ Gyr, respectively. Nevertheless, this timescale should be taken with caution since it is affected by the available snapshots in the simulation.

\section{Discussion and Conclusions} \label{sec:conclusions}

We  statistically analysed  the oxygen abundance gradients of the  star-forming gas in galaxies with a variety of morphologies, selected from \RECAL\, the highest numerical resolution run of the \eagle\  Project \citep{schaye2015}. We studied  the evolution of the metallicity gradients as a function of redshift in relation to the stellar mass, galaxy size, star formation activity and  merger history of the selected galaxies. The merger trees are also followed identifying the progenitors with measured metallicity gradients. 
We highlight that all metallicity gradients  are  estimated within the radial range $[0.5, 1.5]R_{\rm eff}$, in well-resolved systems as explained in Section 2. The values of the gradients depend strongly on the radial range selected to be measured, hence, this issue 
should be considered when comparing our results  with observations and previous numerical works.

Our main results are:
\begin{itemize}
\item We find that the median \sohge\  exhibits a very weak positive redshift evolution of $\sim0.001 $~dex~kpc$^{-1}/\delta z$ (with $r.m.s. = 0.004$) for $z \leq 2$, also consistent with no evolution (Fig.~\ref{gradevol}).
A large diversity of metallicity gradients is detected, which grows with increasing redshift by almost an order of magnitude and is dominated by $\mstar \leq 10^{9.5} \rm M_{\sun}$.  
The  weak positive evolution is confirmed by a positive skewness of the metallicity distributions (Fig.~\ref{histos}). The skewness is only consistent with a more symmetric distribution for normalised gradients at $z \leq 0.5$.
Galaxies with negative metallicity gradients show a redshift evolution of $\sim 0.03$~dex~kpc$^{-1}/\delta z$, becoming shallower for decreasing redshift (Fig.~\ref{diffgradevol}).

\item 
At $z\sim0$, the simulated metallicity gradients  are found to depend  weakly on  galaxy stellar mass  (Table~\ref{table2}). 
This dependence tends to become  slightly stronger with increasing redshift (Fig.~\ref{evolzmasas} and \ref{gradmass}). For $\mstar \sim 10^{9.5} \rm M_{\sun}$, the \sohge\ becomes shallower  by a factor of three  from $z \sim 1.5$ to $z \sim 0$.  However, on one hand,  we note that the {\it r.m.s} associated with the linear fits are relatively high and, hence, this trend should be taken with caution (Table~\ref{table2}). On the other hand, this trend agrees with the weak but statistically significant evolution of the \sohge\  detected for intermediate mass galaxies (see Fig.~\ref{evolzmasas}). For more massive systems the \sohge \ evolution is weaker over the same redshift range. Meanwhile, the \sohge \ of lower stellar mass galaxies tend to become slightly shallower and less disperse  from $z\sim 1$ to $z\sim 0$ (Fig.~\ref{evolzmasas}). These results agree with a narrative where massive galaxies are already more evolved and have converged to a  shallow gradient by $z \sim 1.5$ \citep{molla2020}, whereas galaxies with $\mstar \lsim 10^{10}$M$_\odot$ are still actively forming stars and are more likely to be affected by feedback, particularly those at the lower mass end. As a consequence, their metallicity gradients are still evolving towards shallow values.

\item 
An anti-correlation of negative \sohge\ with the inverse of disc scale-length is detected (Fig.~\ref{gradvsreff}), which  agrees with  observational findings  \citep{bresolin2019} and previous numerical studies \citep[e.g.][]{tissera2019,hemler2020}.  This relation is present since $z \leq 1.5$ and does not  evolve significantly with time. This trend is expected in an inside-out galaxy formation scenario \citep{momaowhite1998} and is consistent with  the fact that these galaxies are more disc-dominated and  their sizes grow systematically with redshift (Fig.\ref{histogradselected}).
Conversely, galaxies with positive \sohge\ determine  a weaker positive relation with galaxy size, reflecting a different evolutionary path,  which is found to be associated with recent merger events that seem to be able to disturb strongly their morphologies, producing more spheroidal-dominated systems.

\item
The merger histories of the {\sc EAGLE} galaxies show that those with strong positive and negative \sohge\ tend to have  more recent major merger events and starbursts than systems with weak \sohge\  (Fig.~\ref{mergertime}).  
Additionally, galaxies with positive  \sohge\  tend to have diluted central metallicity compared to those with  negative ones regardless of stellar mass and redshift (Fig.~\ref{centraloh}). This is the expected behaviour for galaxies that experienced low-metallicity inflows and/or strong metal-loaded outflows \citep[e.g.][]{perez2011, torrey2012}. Galaxies with negative gradients have enhanced central metallicities, indicating continuous star formation activity and/or higher metallicity retained fractions. This suggests that major merger events  could  contribute to induce star formation but do not disturb strongly their morphologies. Complementary, gas reservoirs supplied during these events  could be used to rebuild the disc components more efficiently.

\item
Low and intermediate stellar mass galaxies show a clear anti-correlation of \fgas\ and sSFR with  \taumm\ at all analysed z, indicating that  recent major mergers   tend to be gas-rich and to enhance the star formation activity  in galaxies (Fig.~\ref{tau_prop}).   
These systems show an excess of gas by a factor of two approximately, together with an increase of sSFR by $\sim 0.5$~dex (Fig.~\ref{tau_prop}) with respect to galaxies with  \taumm$ > 4 ~\rm Gyr$. These trends are in agreement with observational results for post merger galaxies for which  an excess of both parameters with respect to the main sequence at $z \sim 0$ has been reported \citep{ellison2018}. Our results  suggest that major mergers may not be able to quench  star formation immediately  in galaxies with $\mstar \leq 10^{9.9} \Msun$. However, more massive systems,  albeit being overall more gas-poor and less star-forming, show a decrease of their gas reservoirs  with decreasing \taumm\,  whilst no clear dependence of  the sSFR on \taumm\ is detected. This suggests that while major mergers  may be efficient triggers of  the transformation of the remaining gas reservoir into stars in massive galaxies, this contribution marginally increase their specific  star formation activity.

\item 
At $z=0$ star-forming \eagle\ galaxies  tend to have weak  \sohge, although their progenitors exhibit a diversity of metallicity slopes as a function of redshift. A fraction of these galaxies evolves from progenitors that tend to  remain with  weak  gradients since $z \sim 2$, and grow in size systematically, becoming more disc-dominated at lower redshift (Fig.~\ref{histogradselected}). Others  are clearly more affected by major mergers that drive changes in the metallicity distributions of their ISM, with a median timescale of $\sim 2$~Gyr. Galaxies with strong negative gradients favour more disc-dominated morphologies, whereas those with strong positive ones  are predominantly more spheroidal.
The median time-scale of variation for the metallicity gradients to evolve from weak to strong and vice-versa is larger than  those corresponding to variations of morphological features  driven by mergers or interactions \citep{lotz2011,bignone2017}. This renders the  possibility of using metallicity gradients to detect galaxies in advanced post merger stages \citep{thorp2019}.

Our results provide insights into the evolution of the metallicity distribution in galaxies encoded by the metallicity gradients in a cosmological context. The comparison with forthcoming observations of galaxies and their chemical abundances at high redshift will allow us to  better understand  the link between chemical patterns and galaxy assembly. 

\end{itemize}

\section*{Acknowledgements}
{We thank the anonymous referee for useful comments that helped to improve this manuscript. 
YRG acknowledges the support of the “Juan de la Cierva Incorporation” fellowship (IJC2019-041131-I). 
PBT acknowledges partial funding by Fondecyt 1200703/2020 (ANID) and ANID Basal Project FB210003. LB acknowledges support from CONICYT FONDECYT POSTDOCTORADO 3180359. 
Part  of this work  was  performed  in Ladgerda Cluster (Fondecyt 1200703/2020).
This project has been supported partially by the European Union Horizon 2020 Research and Innovation Programme under the Marie Sklodowska-Curie grant agreement No 734374.
This work used the DiRAC Data Centric system at Durham University, operated by the Institute for Computational Cosmology on behalf of the STFC DiRAC HPC Facility (www.dirac.ac.uk). This equipment was funded by BIS National E-infrastructure capital grant ST/K00042X/1, STFC capital grants ST/H008519/1 and ST/K00087X/1, STFC DiRAC Operations grant ST/K003267/1 and Durham University. DiRAC is part of the National E-Infrastructure. We acknowledge PRACE for awarding us access to the Curie machine based in France at TGCC, CEA, Bruyeres-le-Chatel.
}

\section*{Data availability}
{The EAGLE simulations are publicly available; see \citet{mcAlpine2016}.}
\bibliographystyle{mnras}
\bibliography{Tissera2021}

\appendix

\section{Comparison between different estimators of the metallicity gradients}
\label{append:grad}

\begin{figure*}
\resizebox{8cm}{!}{\includegraphics[trim={0 30 0 0}]{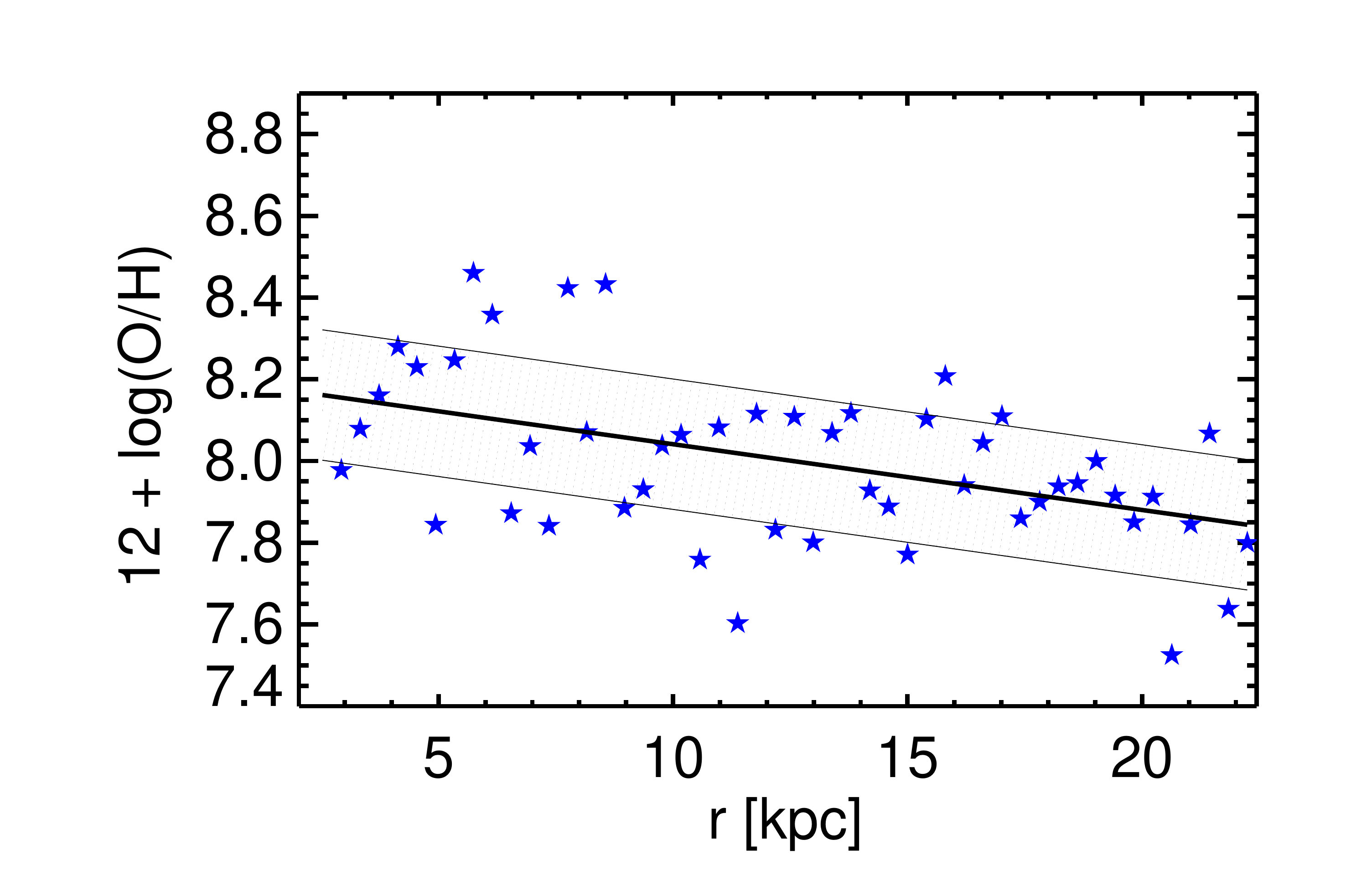}}
\resizebox{8cm}{!}{\includegraphics[trim={0 30 0 0}]{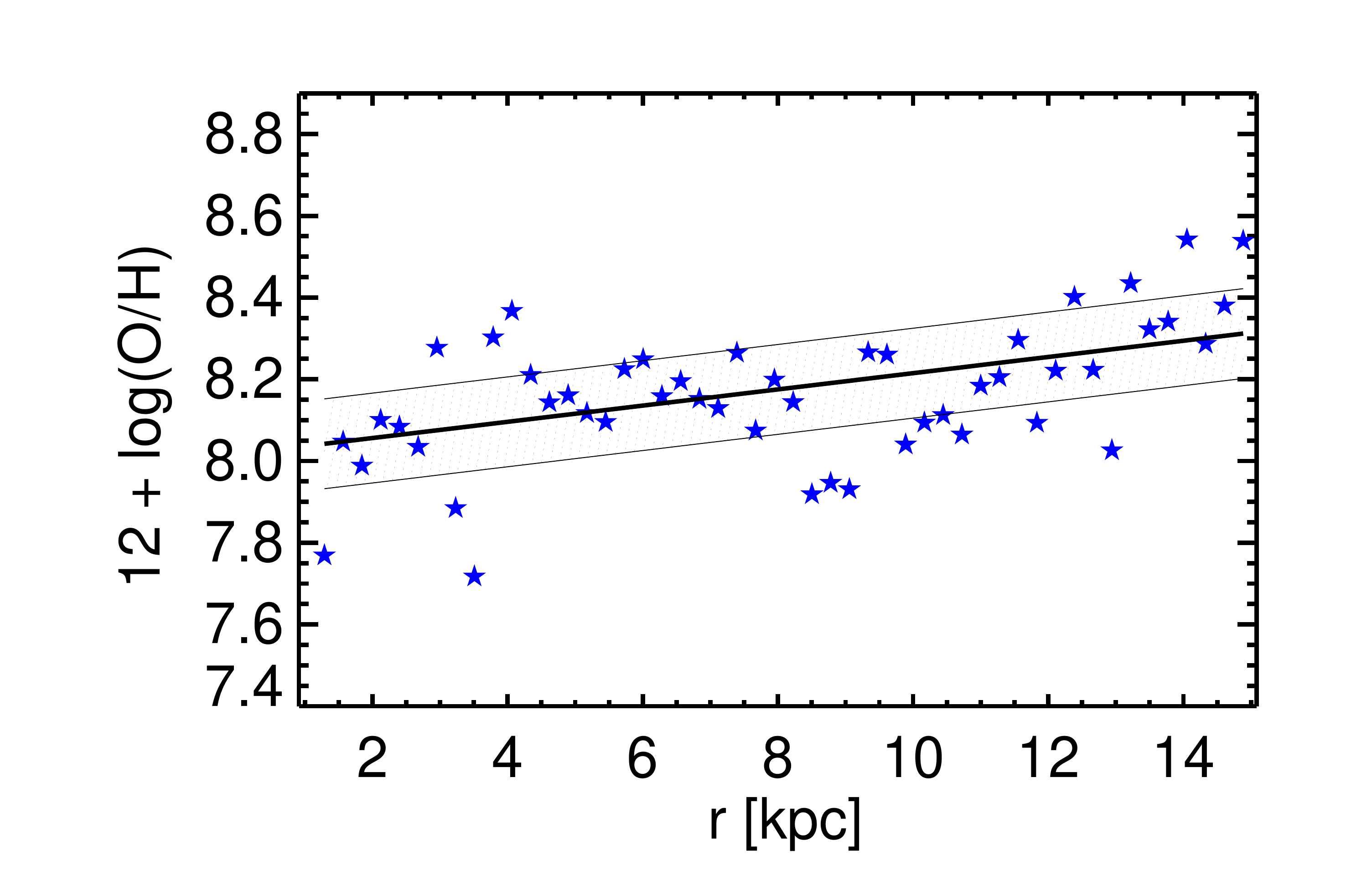}}\\
\resizebox{8cm}{!}{\includegraphics[trim={0 10 0 0}]{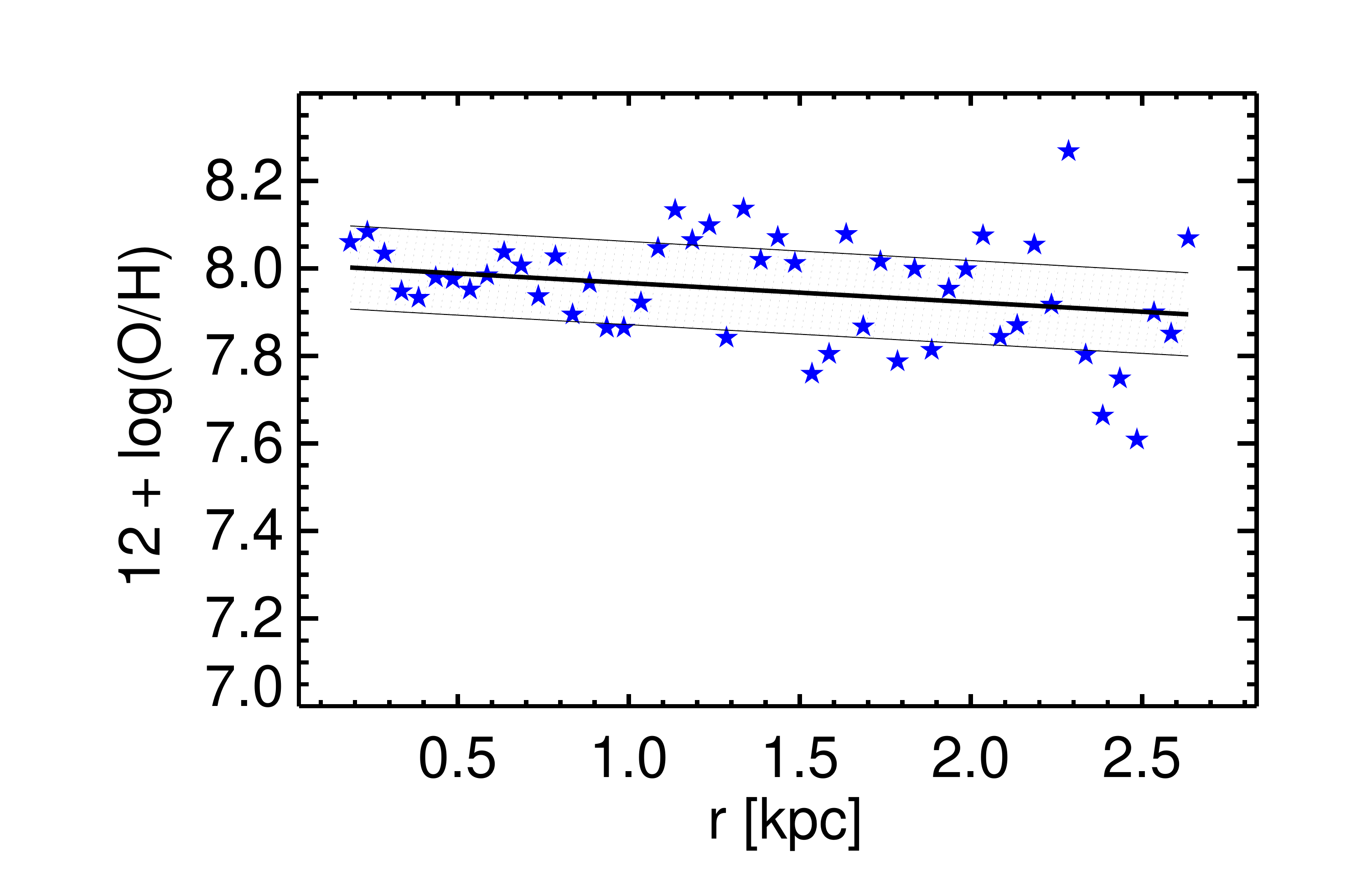}}
\resizebox{8cm}{!}{\includegraphics[trim={0 10 0 0}]{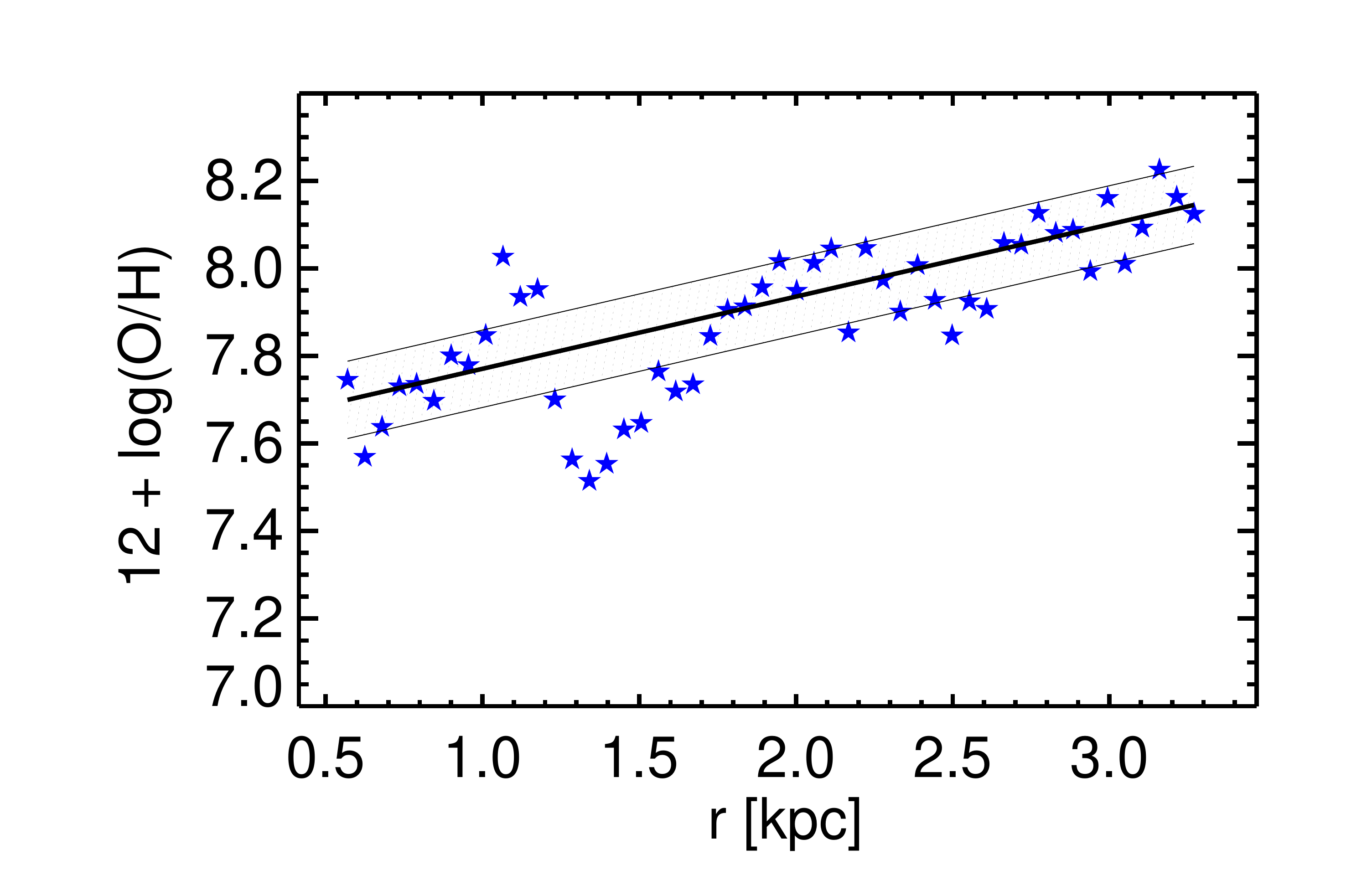}}\\
\caption{Examples of oxygen abundance profiles calculated by using star-forming gas at $z \sim 0$ (upper panels) and $z \sim 2$ (lower panels). The median values estimated per radial interval are displayed (blue stars). The linear regression fits (solid black lines) and their standard deviation (shaded areas) are shown.
}
\label{profilesz0}
\end{figure*}

As discussed in the main section of this paper, a large diversity of metallicity profiles of the SFG components is detected.
The metallicity profiles were constructed by  following the procedure. We used 50 equally-spaced radial bins defined within [0.5, 1.5]\Reff. Only radial bins with more than 10 SFG particles are considered to estimate the medians. 
Then,  a linear regression based on a robust least absolute deviation technique was performed to the metallicity profiles resolved with more than 5 bins with estimated medians, obtaining the slopes (\sohge) or gradients.
Examples of the linear fits to the simulated profiles are shown in Fig.~\ref{profilesz0} for two examples chosen to have similar stellar masses but opposite slopes.
 The median values (blue stars) and linear regressions (black thick lines) are depicted in the radial range $\rm [0.5,1.5]R_{eff}$.  The shaded regions are enclosed by the standard deviations.
We show two galaxies at $z=0$ with similar $\rm \mstar \sim 10^{10.4} \Msun$ and, at $z=2$, with $\rm \mstar \sim 10^{9.7} \Msun$.

\begin{figure*}
\resizebox{18cm}{!}{\includegraphics[trim={0 10 0 0}]{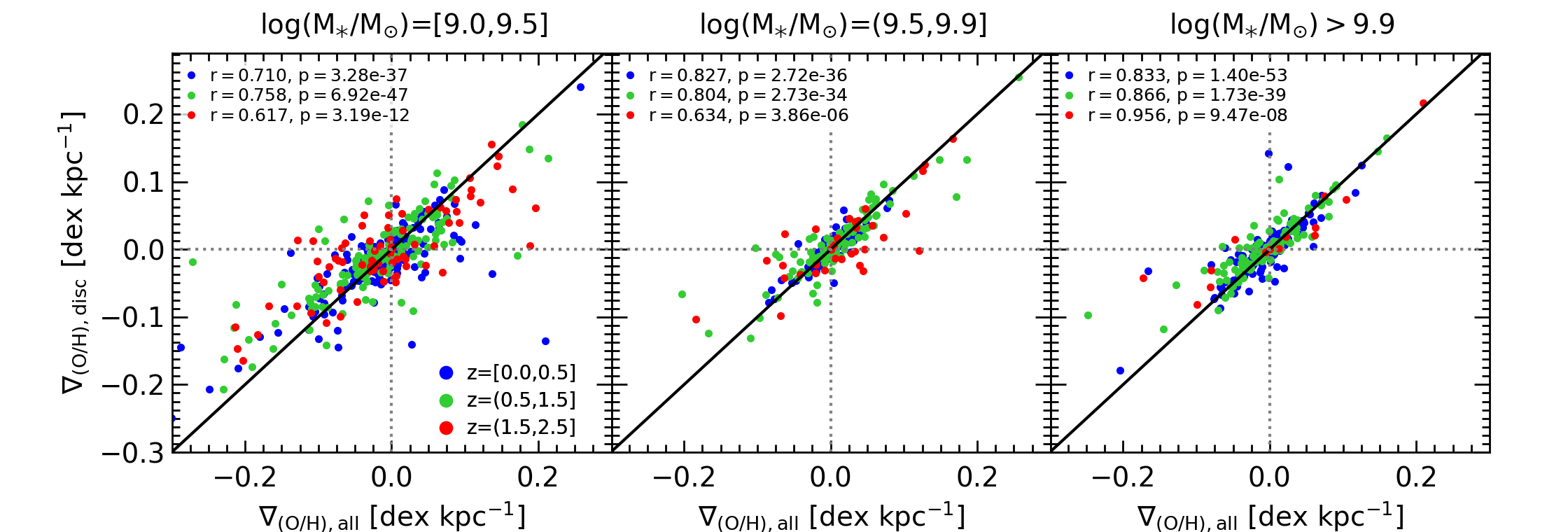}}
\caption{Oxygen abundance slopes computed by using the total SFG ($\rm \nabla_{(O/H),\ all}$) and the SFG in the disc component ($\rm \nabla_{(O/H),\ disc}$), in the selected EAGLE galaxies. The distributions for each mass subsample are shown for three redshift bins: $[0.0,0.5)$ (blue), $[0.5,1.5)$ (green) and $(1.5,2.5]$ (red). The black solid lines denote the 1:1 relation.}
\label{gradgradgas}
\end{figure*}

To test the distribution of SFG and the impact on the metallicity profiles, we calculated them by using  the SFG in the disc components of the selected galaxies ($\sohge_{\rm ,disc}$) and in the whole galaxies ($\sohge_{\rm ,all}$).
In both cases, we fitted a robust linear regression (least absolute deviation - LAD) in the radial range $[0.5,1.5]$\Reff.
In  Fig.~\ref{gradgradgas} we show the correlation between both slopes.
To quantify these trends,  Spearman factors are included in the figure.
As can be seen, the slopes determine an excellent correlation consistent with a 1:1 relation for all stellar masses at all analysed redshifts. The good correlations indicate that most of the SFG within the adopted radial range is actually rotationally supported \citep{trayford2019a}.

 We also tested that when central regions are included the metallicity gradients tend to be more negative than expected, since the gas in the central region will tend to be more chemically enriched  (this is not always the case because some  systems have strong positive gradients and others have broken metallicity profiles). Also, if we extend the radial interval to 50 kpc regardless of the mass of the galaxies, the metallicity gradients are also more negatives since the outer regions are much less enriched \citep[see][for a  discussion on the metallicity gradients in these two regions]{collacchioni2019}. We note that for some galaxies this is not valid since the metallicity profiles show a flatting of the metallicity gradients in the outskirts. These results show that the metallicity profiles are generally not well-fitted by a single linear regression as already reported by observational works \citep[e.g.][]{diaz1989,sanchezmenguiano2016} and, hence, the metallicity gradients depend strongly on the radial range where they are measured. 

\section{Normalised metallicity gradients} \label{append:norm}
\begin{figure*}
\resizebox{5.5cm}{!}{\includegraphics[trim={0 20 0 0}]{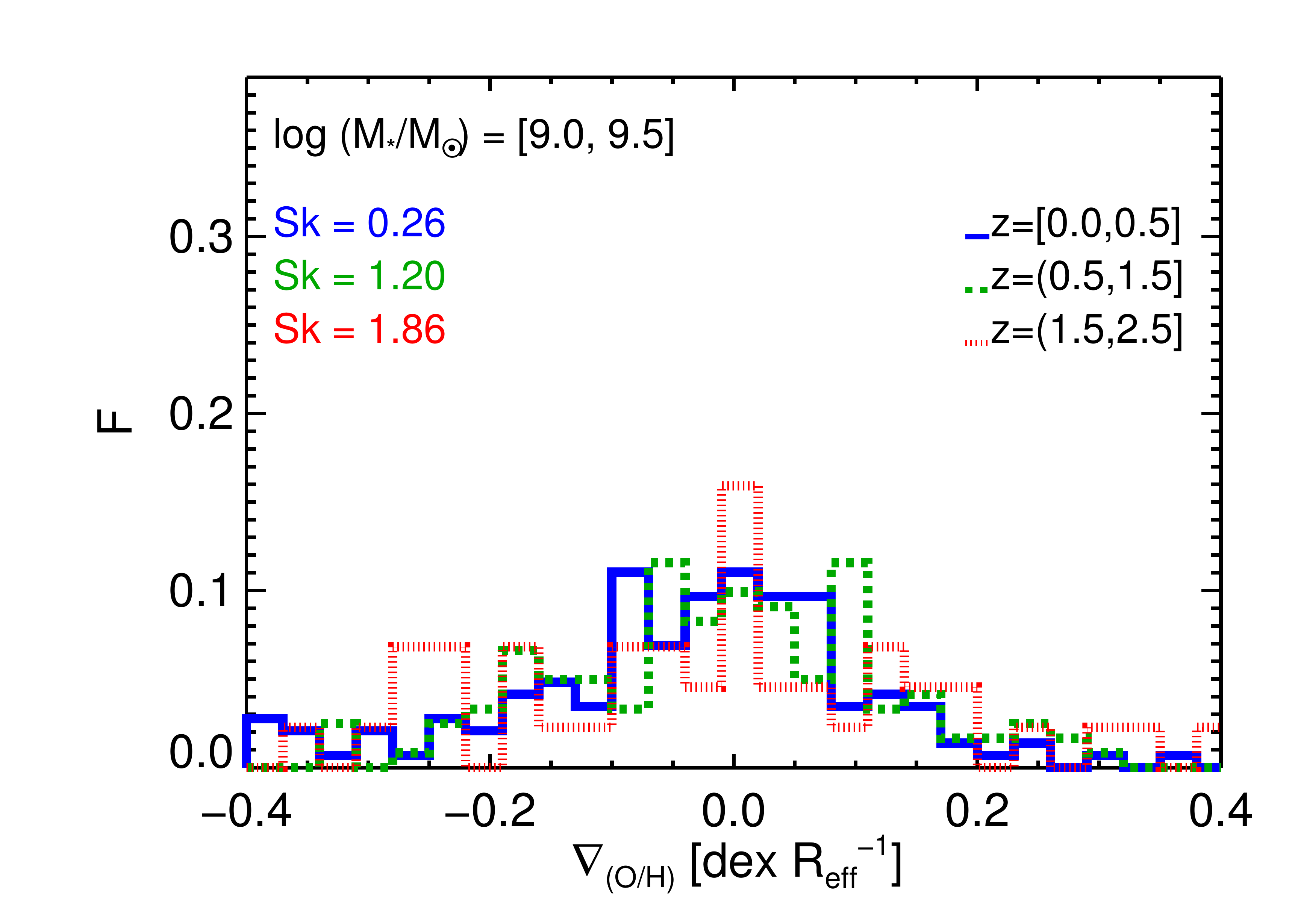}}
\resizebox{5.5cm}{!}{\includegraphics[trim={0 20 0 0}]{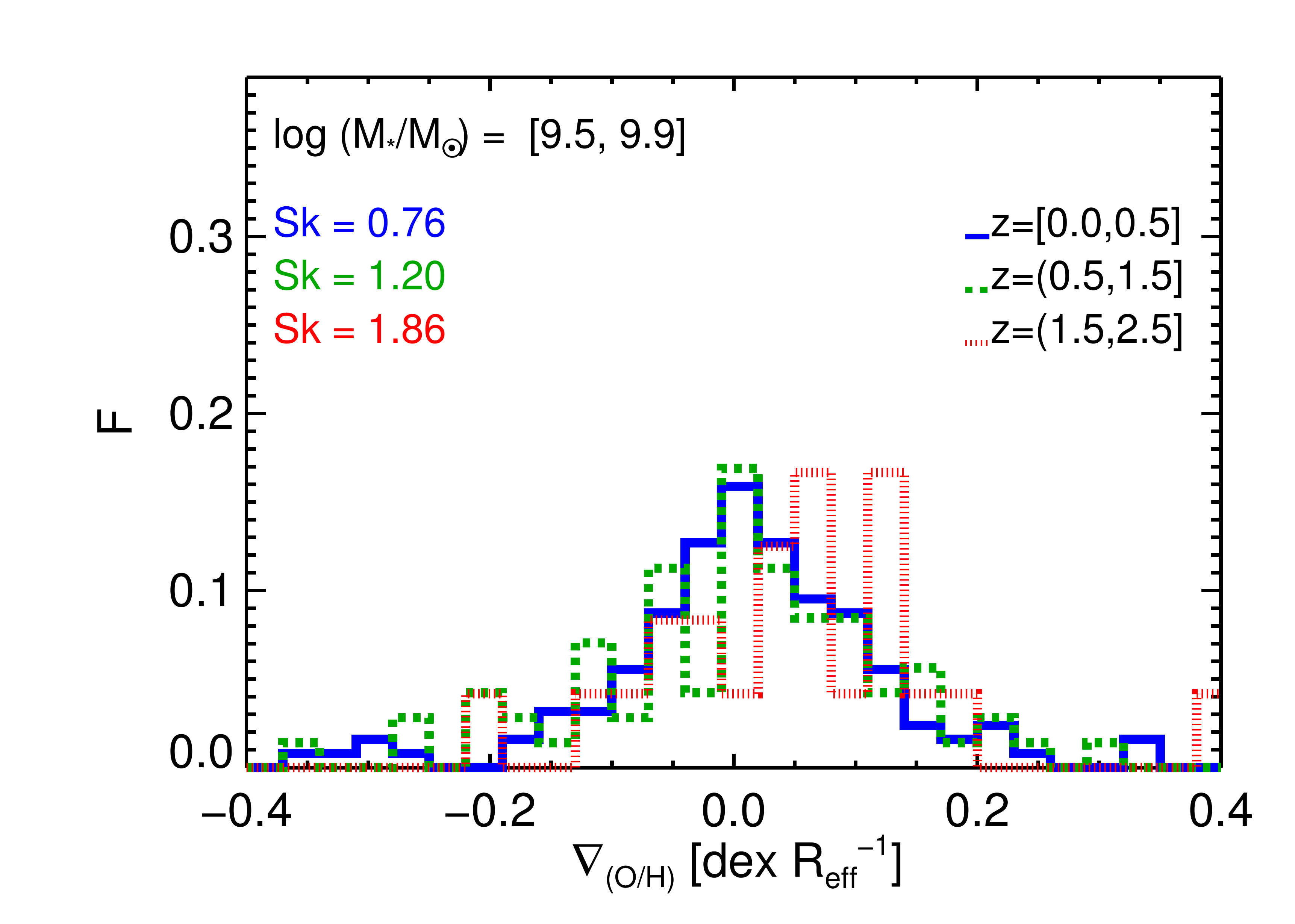}}
\resizebox{5.5cm}{!}{\includegraphics[trim={0 20 0 0}]{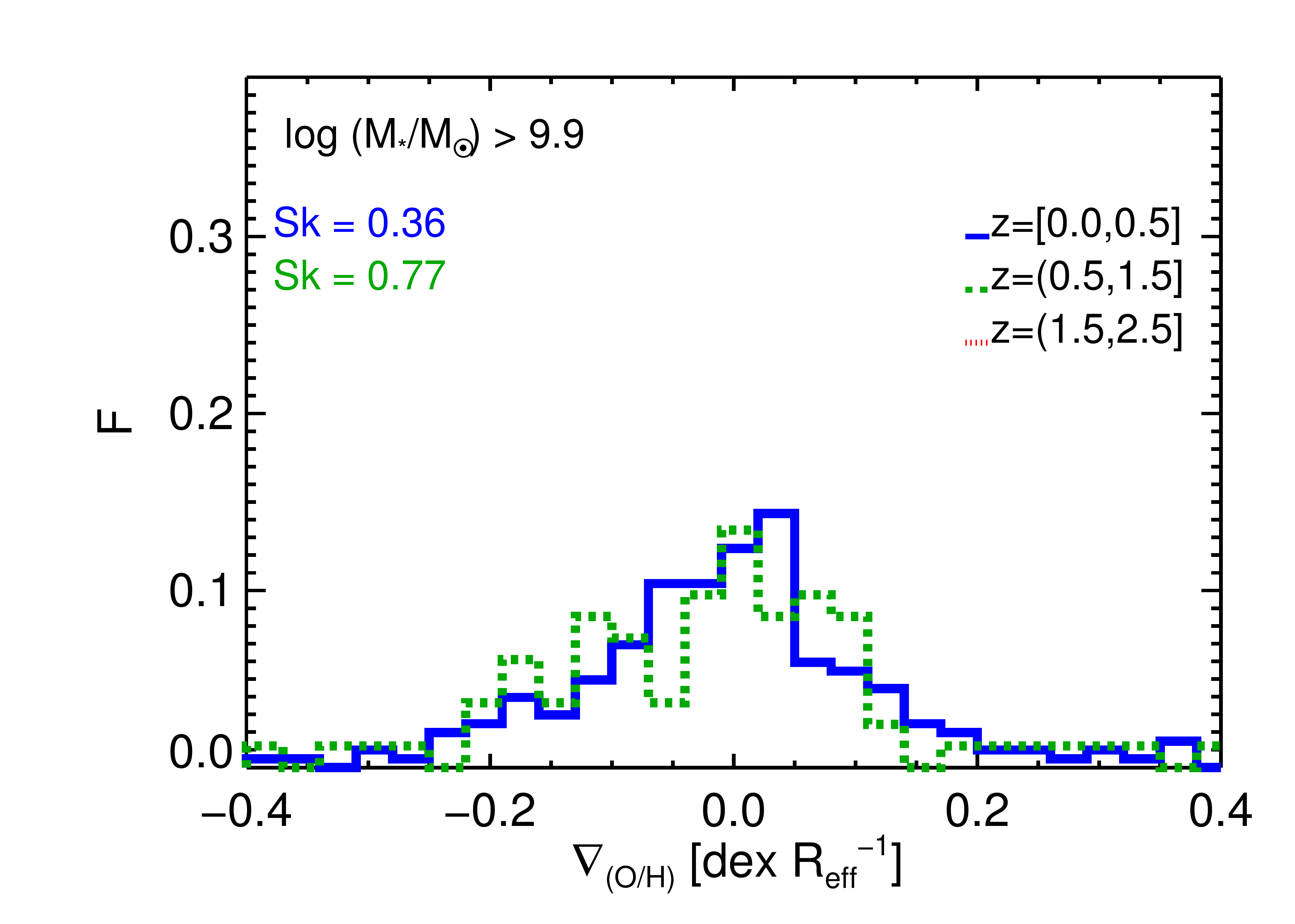}}
\caption{Distributions of normalised \sohge\ calculated for SFG
of the selected \eagle\ galaxies. The distributions for each of three stellar mass ranges are displayed for three redshift intervals: $[0.0,0.5]$ (blue, solid  lines), $(0.5,1.5]$ (green, dashed  lines) and $(1.5,2.5]$ (red, dotted lines).
}
\label{histosnorm}
\end{figure*}

We also analysed the distribution of \sohge\ 
normalised by the scale-lengths of the galaxy.
In agreement with observations \citep{sanchez2013Califa,kewley2010}, the distributions became wider as shown in Fig.~\ref{histosnorm}. We estimate the skewness for each of them, finding that $Sk < 1$, within the redshift range $[0,0.5]$.
For higher redshift, we measure $Sk \ge 1$  with no systematic trend. Overall we found that the tail towards positive gradients detected in Fig.~\ref{histos} is erased by normalising the gradients at $z \leq 0.5$. 

However, we decided to work with the metallicity gradients in \dexkpc for two reasons. First, the determination of sizes is quite complicated at high redshift and most of the available data is in \dexkpc.  Secondly, the\RECAL~simulation exhibits a slight downturn in the mass-size relations for massive galaxies for $z > 0.5$ \citep[see][]{furlong2017}.


\section{Dependence of galaxy properties on \taustar} \label{append:sec_25}

\begin{figure*}
\resizebox{16cm}{!}{\includegraphics[trim={0 0 0 0}]{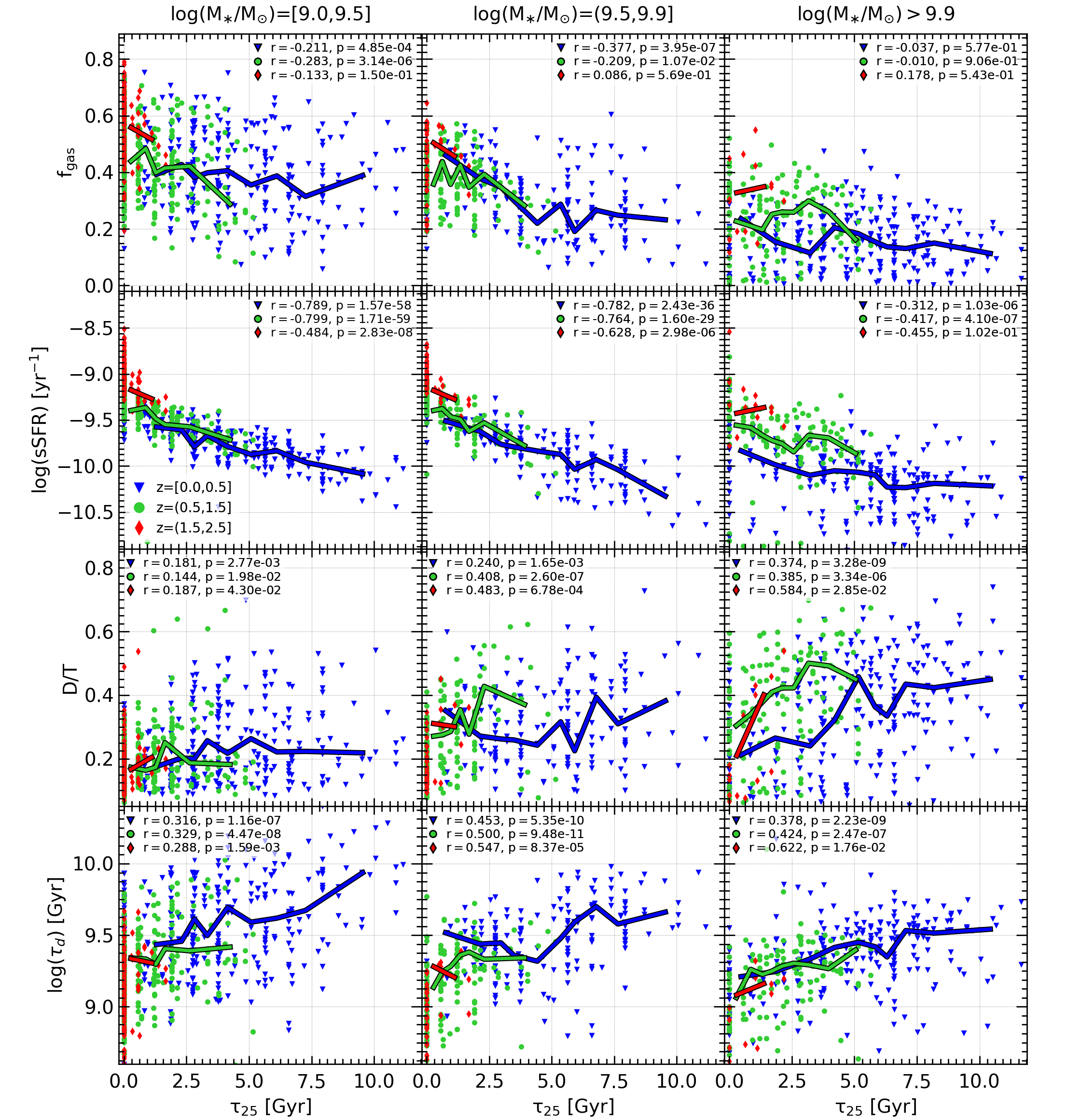}}
\caption{Distributions of \fgas, sSFR,  D/T and $\tau_{\rm d}$ as a function of \taustar ~for galaxies with low, intermediate and high stellar masses (left, middle, and right panels). In each panel, the distributions (symbols) and the medians (solid lines) are distinguished according the three defined redshift intervals (indicated in the inset labels). The Spearman correlations factors are also included.
}
\label{tau25prop}
\end{figure*}

Here we show the dependence of \fgas, sSFR, D/T and $\tau_{\rm d}$ as a function of \taustar. As explained in Section~\ref{sec:mergers}, this time provides an estimation of the occurrence of the last major increase of $\mstar$ in a galaxy. It could be due to major merger or to other mechanisms (minor merger, interactions, inner instabilities), which will not be analysed in detail in this paper. The fact that \taustar ~is agnostic to the specific origin of the sudden increase of stellar mass, it provides an estimation of their combined effects.
Figure~\ref{tau25prop} displays the relation between galaxies properties as a function of \taustar ~in a similar way to Fig.~\ref{tau_prop}.
The trends are consistent with those found as a function of \taumm. 
In the case of the correlation with sSFR because both parameters, the anticorrelation is expected by construction.

\end{document}